\documentclass[useAMS,usenatbib]{mn2e}
\usepackage{times,mathptm}
\usepackage{lscape}
\usepackage{psfig}
\usepackage{subfigure}

\def\lbsol{$L_{B,\odot}$}

\def\chandra{{\it Chandra\/}}
\def\xray{\hbox{X-ray}}

\def\gs{\mathrel{\raise0.35ex\hbox{$\scriptstyle >$}\kern-0.6em\lower0.40ex\hbox{{$\scriptstyle \sim$}}}}
\def\ls{\mathrel{\raise0.35ex\hbox{$\scriptstyle <$}\kern-0.6em\lower0.40ex\hbox{{$\scriptstyle \sim$}}}}

\def\Wm2{\,\hbox{W}\,\hbox{m}^{-2}}
\def\gsim{\mathrel{\raise0.35ex\hbox{$\scriptstyle >$}\kern-0.6em\lower0.40ex\hbox{{$\scriptstyle \sim$}}}}
\def\lsim{\mathrel{\raise0.35ex\hbox{$\scriptstyle <$}\kern-0.6em\lower0.40ex\hbox{{$\scriptstyle \sim$}}}}
\def\ltsima{$\; \buildrel < \over \sim \;$}
\def\simlt{\lower.5ex\hbox{\ltsima}}
\def\gtsima{$\; \buildrel > \over \sim \;$}
\def\simgt{\lower.5ex\hbox{\gtsima}}

\usepackage{natbib}
\usepackage{graphicx}
\usepackage{floatflt}
\usepackage{amssymb}

\newcommand{\ergcms}{erg~cm$^{-2}$~s$^{-1}$}
\newcommand{\ergs}{erg~s$^{-1}$}

\newcommand{\etal}{et\,al.}

\newcommand{\Nh}{$N_{\rm H}$}

\newcommand{\comment}[1]{}

\newcommand{\Spitzer}{\textit{Spitzer}}

\newcommand{\Chandra}{\textit{Chandra}}

\title[Hot Gas and Radio AGN Activity in ETGs]{The Cosmic History of Hot Gas Cooling and Radio AGN Activity in Massive Early-Type Galaxies}
\author[Danielson et al.]
{\parbox[h]{\textwidth}{
A.\,L.\,R.\ Danielson,$^{1}$
B.\,D.\ Lehmer,$^{1,2,3}$
D.\,M.\ Alexander,$^{1}$
W.\,N.\ Brandt,$^{4,5}$
B.\,Luo,$^6$
N.~Miller,$^7$
Y.\,Q.\ Xue$^{4,5}$
\& J.\,P.\ Stott$^{1}$ 
}
\vspace*{6pt}\\
 \\ 
$^{1}$Department of Physics, Durham University, South Road, Durham, DH1 3LE,
U.K.\\ 
$^{2}$The John Hopkins University, Homewood Campus, Baltimore, MD 21218, USA\\ 
$^{3}$NASA Goddard Space Flight Centre, Code 662, Greenbelt, MD 20771, USA\\
$^{4}$Department of Astronomy and Astrophysics, Pennsylvania State University, University Park, PA 16802, USA\\
$^5$Institute for Gravitation and the Cosmos, Pennsylvania State University,
University Park, PA 16802, USA\\
$^6$Harvard-Smithsonian Center for Astrophysics, 60 Garden Street, Cambridge,
MA 02138, USA\\
$^7$Department of Astronomy, University of Maryland, College Park, MD,
20742-2421, USA\\
}

\begin{document}

\date{\today}

\pagerange{\pageref{firstpage}--\pageref{lastpage}} \pubyear{2011}

\maketitle

%
\begin{abstract} 
%

We study the X-ray properties of 393 optically selected early-type galaxies
(ETGs) over the redshift range of $z\approx$~0.0--1.2 in the {\it Chandra} Deep
Fields.   To measure the average X-ray properties of the ETG population, we
use X-ray stacking analyses with a subset of 158 passive ETGs (148 of which
were individually undetected in X-ray).  This ETG subset was constructed to
span the redshift ranges of $z =$~0.1--1.2 in the $\approx$4~Ms CDF-S and
$\approx$2~Ms CDF-N and $z =$~0.1--0.6 in the $\approx$250~ks E-CDF-S where the
contribution from individually undetected AGNs is expected to be negligible in
our stacking. We find that 55 of the ETGs are detected individually in the
X-rays, and 12 of these galaxies have properties consistent with being passive
hot-gas dominated systems (i.e.,\ systems not dominated by an X-ray bright
Active Galactic Nucleus; AGN). On the basis of our analyses, we find little
evolution in the mean 0.5--2~keV to $B$-band luminosity ratio ($L_{\rm X}/L_{B}
\propto [1+z]^{1.2}$) since $z \approx 1.2$, implying that some heating
mechanism prevents the gas from cooling in these systems.  We consider that
feedback from radio-mode AGN activity could be responsible for heating the gas.
We select radio AGNs in the ETG population using their far-infrared/radio flux
ratio. Our radio observations allow us to constrain the duty cycle history of
radio AGN activity in our ETG sample.  We estimate that if scaling relations
between radio and mechanical power hold out to $z \approx 1.2$ for the ETG
population being studied here, the average mechanical power from AGN activity
is a factor of $\approx$1.4--2.6 times larger than the average radiative
cooling power from hot gas over the redshift range $z \approx$~0--1.2.  The
excess of inferred AGN mechanical power from these ETGs is consistent with that
found in the local Universe for similar types of galaxies. \\

\end{abstract}


\section{Introduction}

\label{Intro} 
The most successful theoretical models of galaxy evolution
(e.g. \citealt{Bower06} and \citealt{Croton06}) require that feedback,
in the form of energetic outflows from active galactic nuclei (AGNs),
will have a fundamental influence on the evolution of intermediate and
massive galaxies.  In these models, energy injected from AGN radio
jets heats the interstellar and intergalactic mediums of massive
early-type galaxies (ETGs) and further drives interstellar gas out of
these systems.  This energy injection effectively quenches star
formation and supermassive black hole (SMBH) accretion and prevents
galaxies and SMBHs growing.  The interstellar gas itself is thought to
be produced by evolving stars ejecting material through stellar winds
and supernovae (at a rate of $\approx
1.3[L_{B}/10^{11}L_{B,\odot}]M_{\odot} yr^{-1}$; e.g.,
\citealt{Mathews03B} and \citealt{Bregman09}), as well as gas infall
from the intergalactic medium.

The hot gas in massive ETGs ($\simgt$$10^{11} M_\odot$) has been found
to radiate powerfully at \xray\ wavelengths through thermal
bremsstrahlung and yet it does not appear to be cooling as expected.
The \xray\ spectral energy distributions (SEDs) of massive ETGs
demonstrate that hot gas ($kT =$~0.3--1~keV) typically dominates the
\hbox{0.5--2~keV} emission (e.g., \citealt{Boroson11}).  The
temperatures and densities in the central regions imply relatively
short radiative cooling times of $\approx10^8$~yr
(\citealt{Mathews03B}).  However, large quantities of cool
($10^{4-5}$K) gas are not observed in the ETGs, which would be
predicted by simple cooling flow models (see, e.g.,
\citealt{Mathews03B} for a review).  By these observational arguments,
it is necessary that some feedback mechanism (e.g., the AGN radio jets
predicted by the models) keeps the gas hot and/or expels the cooled
gas reservoirs.

Direct observational evidence for the interaction of AGN radio jets
with hot gas has been obtained via \xray\ and radio observations of
massive ETGs in the local universe (e.g., \citealt{Boehringer93};
\citealt{Birzan04}; \citealt{Forman05}; \citealt{Rafferty06}). These
observations have revealed relativistic radio outflows inflating large
\xray-emitting gas cavities with cool gas observed at the cavity rims.
Measurements of \xray\ cavity sizes and their surrounding gas
densities and temperatures can give estimates of the mechanical energy
input required by the radio jets to inflate the cavities against the
pressure of the surrounding gas.  The derived mechanical energy and
jet power are in the ranges $10^{55}$--$10^{62}$~ergs and
$10^{40}$--$10^{46}$~erg~s$^{-1}$, respectively (\citealt{McNamara09},
Nulsen et al 2007; astro-ph/0611136), sufficient to suppress gas
cooling in the galaxy and impede star formation and cold gas SMBH
accretion (\citealt{Allen06}).  This type of feedback has been
referred to as `radio mode' or `maintenance mode,' since during this
phase the accretion rate onto the central SMBH driving the AGN is low,
and thus the majority of the gas heating is through radio AGN
activity.

These previous studies clearly indicate that heating by radio jets is
an important process in galaxy evolution. Investigations of the
influence of radio AGN on gas cooling in the general ETG population
(\citealt{Best05}; hereafter B05) found that more than 30\% of the
most massive ($\sim 5 \times 10^{11}$M$_{\odot}$) galaxies host a
radio AGN, which is consistent with radio AGN activity being eposidic
with duty cycles of \hbox{$\approx$$10^{7}$--$10^{8}$~yr}
\citep{Best06}.  Evidence of such episodic radio luminous activity is
also implied by the presence of multiple bright rims and shocks in the
X-ray and radio images of individual ETGs (e.g., M87;
\citealt{Forman05}).  Therefore the prevention of the cooling of large
quantities of gas is thought to be maintained by a self-regulating AGN
feedback loop.  Cooling gas in the ETG centre initially provides a
slow deposition of fuel for SMBH accretion, which in a
radiatively-inefficient accretion mode, leads to the production of a
radio outburst.  Surrounding cool gas is physically uplifted by the
radio outbursts, which increases the gravitational potential energy of
the gas or removes it from the system entirely
(e.g. \citealt{Giodini10}).  As the gas further cools via
\xray\ emission, it falls back towards the SMBH where it can re-ignite
a new cycle of accretion, thus completing the feedback loop
(\citealt{Best06}; \citealt{McNamara07}).  The importance of the role
of feedback from moderately radio luminous AGN is becoming
increasingly apparent, since the feedback energy can be directly
diffused into the interstellar medium \citep{Smolcic09}.

A more complete understanding of the history of gas cooling and
feedback heating in the massive ETG population requires direct
\xray\ and radio observations, respectively, of distant ETG
populations covering a significant fraction of cosmic history.  At
present, such studies are difficult due to the very deep
\xray\ observations required to detect the hot \xray\ emitting gas in
such distant populations (however, see, e.g., \citealt{Ptak07} and
\citealt{Tzan08} for some early work).  Notably, \cite{Lehmer07}
utilised \xray\ stacking techniques and the $\approx$250~ks Extended
$\Chandra$ Deep Field South (\hbox{E-CDF-S}) and $\approx$1~Ms
\chandra\ Deep Field-South \hbox{CDF-S} to constrain the evolution of
hot gas cooling (via soft \xray\ emission) in optically luminous ($L_B
\simgt 10^{10}$~\lbsol) ETGs over the redshift range of \hbox{$z
  \approx$~0--0.7}.  This study showed that the mean \xray\ power
output from optically luminous ETGs at $z \approx 0.7$ is
$\approx$1--2 times that of similar ETGs in the local universe,
suggesting the evolution of the hot gas cooling rate over the last
$\approx$6.3~Gyr is modest at best.  Considering the relatively short
gas cooling timescales for such ETGs ($\sim$$10^8$~yr), this study
provided indirect evidence for the presence of a heating source.
Lehmer \etal\ (2007) found rapid redshift evolution for
\xray\ luminous AGNs in the optically luminous ETG population, which
given the very modest evolution of the hot gas cooling, suggests that
AGN feedback from the radiatively-efficient accreting SMBH population
is unlikely to be the mechanism providing significant feedback to keep
the gas hot over the last $\approx$6.3~Gyr.  However, the mechanical
feedback from radio AGNs, which is thought to be an important AGN
feedback component, was not measured.

In this paper, we improve upon the \cite{Lehmer07} results in the
following key ways: (1) we utilise significantly improved
spectroscopic and multiwavelength photometric data sets to select hot
gas dominated optically luminous ETGs (via rest-frame colours,
morphologies, and spectroscopic/photometric redshifts) and sensitively
identify AGN and star-formation activity in the population (see $\S$~2
and 3); (2) we make use of a larger collection of \chandra\ survey data
(totaling a factor of $\approx$4 times the \chandra\ investment used
by Lehmer \etal\ 2007) from the $\approx$2~Ms \chandra\ Deep
Field-North (CDF-N; Alexander \etal\ 2003), the $\approx$4~Ms CDF-S
(\citealt{Xue11}), and the $\approx$250~ks E-CDF-S (Lehmer
\etal\ 2005) (collectively the CDFs), which allows us to study the
properties of hot gas (e.g., luminosity and temperature) in optically
luminous ETGs to $z \approx 1.2$; and (3) we make use of new radio
data from the VLA to measure the radio luminous AGN activity and the
evolution of its duty cycle in the ETG population and provide direct
constraints on the radio jet power available for feedback.  The paper
is organised as follows.  In $\S$~2, we define our initial working
sample and discuss the ancillary multiwavelength data used to identify
non-passive ETG populations. In $\S$~3, we use various selection
criteria to identify passive ETGs and ETGs hosting radio AGNs.  In
$\S$~4, we constrain the evolution of the X-ray emission from hot gas
in our passive ETG sample using \xray\ stacking techniques.  In
$\S$~5, we discuss the level by which radio AGN can provide heating to
the hot gas in the ETG population.  Finally, in $\S$~6, we summarize
our results.  Throughout this paper, we make use of Galactic column
densities of \Nh$ = 1.6\times10^{20}$~cm$^{-2}$ for the CDF-N
(Lockman~2004) and \Nh$ = 8.8\times10^{19}$~cm$^{-2}$ for the E-CDF-S
region (which also includes the CDF-S; \citealt{Stark92}).  In our
\xray\ analyses, we make use of photometry from 5 bands: the full band
(FB; 0.5-8 keV), soft band (SB; 0.5-2 keV), soft sub-band 1 (SB1;
0.5-1 keV), soft sub-band 2 (SB2; 1-2 keV) and hard band (HB; 2-8
keV).  The following constants have been assumed, $\Omega_{\rm
  M}=0.3$, $\Omega_{\rm \Lambda}=0.7$ and
$H_{0}=70$~km~s$^{-1}$Mpc$^{-1}$ implying a lookback time of 8.4~Gyr
at $z = 1.2$.  Throughout the paper, optical luminosity in the
$B$-Band ($L_{\rm B}$) is quoted in units of $B$-band solar luminosity
($L_{\rm B,\odot} = 5.2\times10^{32}$~erg~s$^{-1}$).

\section{Early-Type Galaxy Sample Selection}
\label{sample}
%

\begin{figure*}
\centerline{\psfig{figure=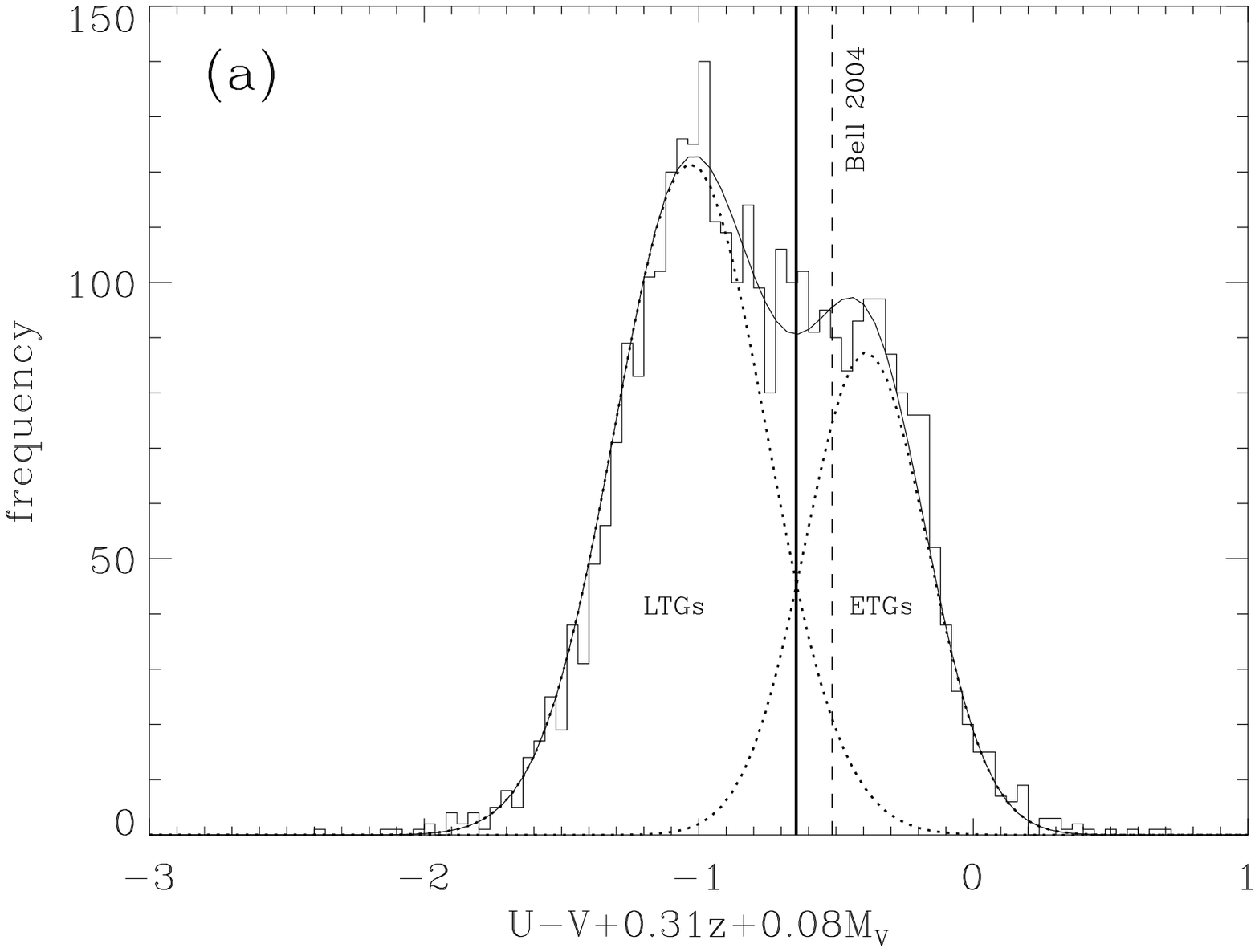,width=3.7in,height=2.7in}
\psfig{figure=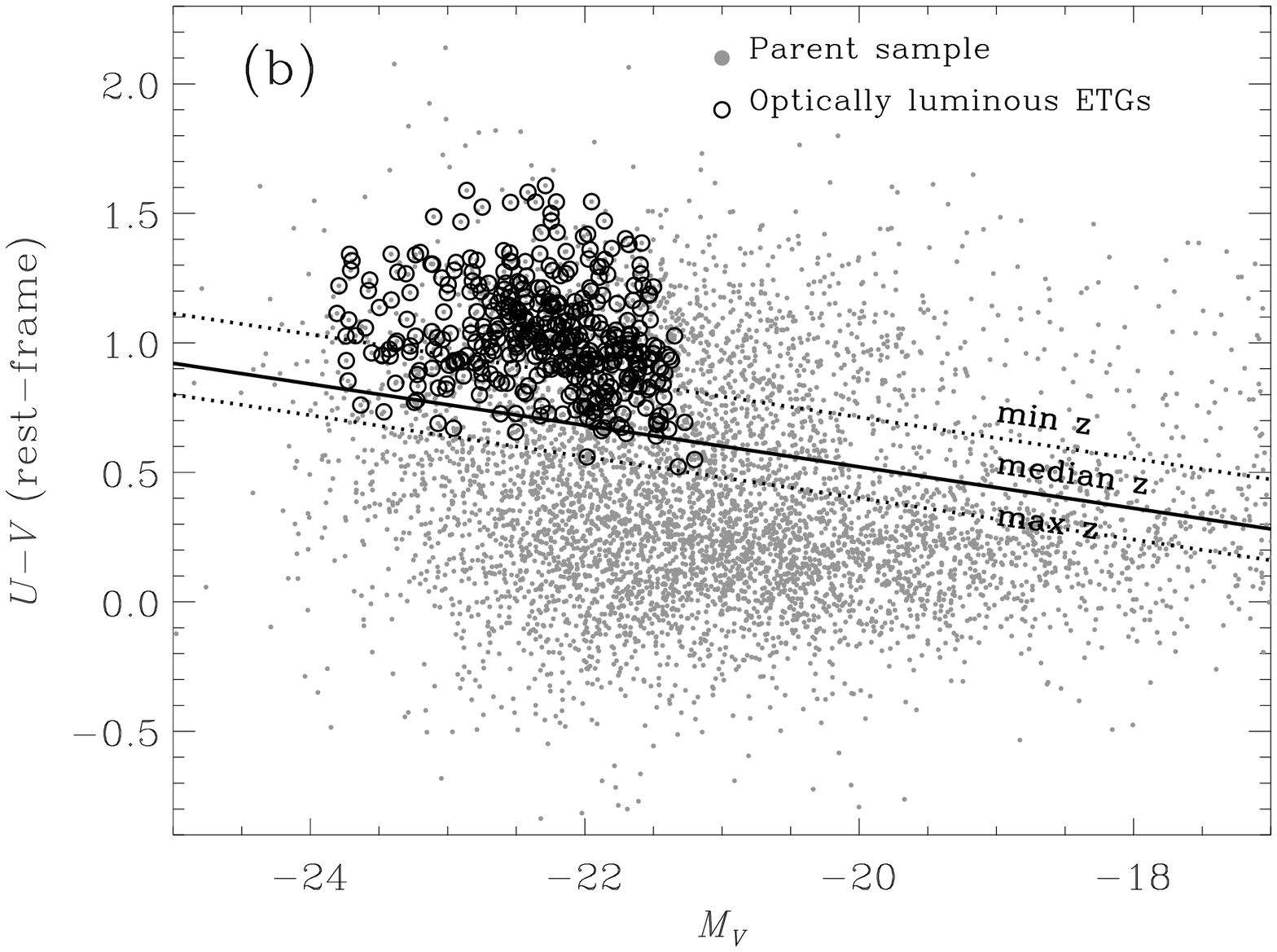,width=3.7in,height=2.7in}}
\caption{(a): Histogram of the entire sample of sources after imposing the
selection criteria (i) to (iii) in $\S$ \ref{sample}, showing the bimodal
distribution of late-type and early-type galaxies. The dashed line shows the
\protect\cite{Bell04} colour-cut, whereas the solid line shows the cut that we
apply, slightly offset due to our different selection methods, using the
redshift dependency taken from \protect\cite{Bell04} but varying the constant.
The dotted and solid curves correspond to fitting a double gaussian to the
sample in order to separate out the late and early-type populations, where the
solid curve is the combination of the two gaussians. (b): Rest frame $U-V$
colour versus absolute (Vega) $V$-band magnitude. The small filled grey circles
represent the whole orginal sample of sources, again after imposing the
selection criteria (i) to (iii) in $\S$ \ref{sample} (including both ETGs
and LTGs). The open symbols represent the 393 sources in our final sample of
ETGs and the dotted lines represent the colour cut applied using the minimum
source redshift and maximum source redshift in our sample, with the line for
the median redshift shown as a solid line (this cut was then applied to our
model data later in Fig.  \ref{fig:IRAC}).}
\label{fig:colour_plot}
\end{figure*}

The primary goals of this study are to constrain the potential heating
from AGN activity and the cooling of the hot gas in the optically
luminous (massive) ETG population over the redshift
range \hbox{$z =$~0.0--1.2} (i.e., over the last 8.4~Gyr).  To
achieve these goals optimally, we constructed samples of optically
luminous ETGs in the most sensitive regions of the
CDFs.

We began our galaxy selection using master optical source catalogues
in the CDF-N and E-CDF-S, which contain a collection of
IR--to--optical photometric data and good redshift estimates (either
spectroscopic redshifts or photometric redshifts).  The CDF-N master
source catalogue consists of 48,858 optical sources detected across
the entire CDF-N region (see \citealt{Rafferty11}).  This catalogue is
based on optical sources detected in the Hawaii HDF-N optical and
near-IR catalogue from \cite{Capak04}, and includes cross-matched
photometry from GOODS-N through $HST$ ACS and $Spitzer$ IRAC
catalogues (e.g., \citealt{Giavalisco04}), $GALEX$
photometry,\footnote{see http://galex.stsci.edu/GR4/.} and deep
$K_s$-band imaging \citep{Barger08}.  In the E-CDF-S, we made use of a
master catalogue of 100,318 sources (see Rafferty et~al.\ 2011).  This
catalogue is based on the MUSYC (\citealt{Gawiser06}), COMBO-17
(\citealt{Wolf04}), and the \hbox{GOODS-S} (\citealt{Grazian06})
optical surveys, and includes cross-matched photometry from MUSYC
near-IR (\citealt{Taylor09}), SIMPLE $Spitzer$ IRAC (Damen
\etal\ 2011), $GALEX$ (see footnote~1), and \hbox{GOODS-S} deep
$U$-band photometry (\citealt{Nonino09}). Our master catalogs are
estimated to be complete to R$\le26$ (see section 2.1 of \citealt{Xue10}).

Whenever possible, we utilised secure spectroscopic redshifts, which
were collected from a variety of sources in the literature and
incorporated into the master source catalogues discussed
above.\footnote{For a comprehensive list of spectroscopic references,
  see Rafferty et~al.\ (2011).}  When spectroscopic redshifts were not
available, we made use of high-quality photometric redshifts, which
were calculated by Rafferty et~al.\ (2011) using an extensive library
of spectral templates (appropriate for galaxies, AGNs, hybrid galaxy
and AGN sources, and stars), the optical--to--near-IR photometry
discussed above, and the Zurich Extragalactic Bayesian Redshift
Analyzer (ZEBRA; \citealt{Feldmann06}). We compared these redshifts to
the photometric redshift catalogue of \cite{Cardamone10} finding a
median difference of 0.01$\pm0.16$ between $z=0.0-0.8$ and
0.01$\pm0.32$ between $z=0.8-1.2$ in the two catalogues, thus
providing additional evidence for the validity of these redshifts.

Starting with the master catalogues of 149,176 collective CDF sources,
we imposed a series of selection criteria that led to the creation of
our optically luminous ETG catalogue that we use throughout this
paper; the imposed selection criteria are summarized below:

\begin{enumerate}

\item We restricted ETG catalogue inclusion to sources with $HST$
  optical magnitudes of $z_{850} < 23$ that were measured to be
  cosmologically distant (i.e., $z > 0.05$). The requirement of
  $z_{850} < 23$ ensures that the photometric redshifts of the
  remaining sources are of high quality \footnote{ CDFN: median
    \hbox{$\mid z_{\rm spec} - z_{\rm phot}\mid/(1 + z_{\rm spec})
      \approx$} 0.015, mean$\approx$0.032 and
    dispersion$\approx$0.090; E-CDF-S: median \hbox{$\mid z_{\rm spec}
      - z_{\rm phot}\mid/(1 + z_{\rm spec}) \approx$} 0.007,
    mean$\approx$0.016 and dispersion$\approx$0.046, for $z_{850} <
    23$ sources.} and provides a highly optically complete (see
  Fig.~2) sample of relatively bright optically luminous ETGs out to
  $z \approx 1.2$. Note that these photo-zs were computed using a
  redshift training procedure that implements spectroscopic redshifts.
  The true accuracy of the photometric redshifts is expected to be $<$
  6-7 times worse than those available for sources with spectroscopic
  redshifts (see Luo et al. 2010 for details). This requirement
  further restricts our study to sources where $HST$ imaging is
  available, thus allowing for further visual inspection of the
  optical morphologies to reasonably good precision (see criterion~v
  below).  This criterion restricted our working sample to 9732 CDF
  sources.

\item We required that the sources are located within 6\arcmin\ of at
  least one of the six independent $Chandra$ aimpoints in the CDFs
  (i.e., the $\approx$4~Ms CDF-S, four $\approx$250~ks E-CDF-S
  pointings, and $\approx$2~Ms CDF-N).  This criterion ensures that
  the galaxies are located in regions where the $Chandra$ imaging is
  most sensitive and of highest quality (e.g., in these regions the
  $Chandra$ point-spread function is small and relatively symmetric).
  Applying this additional restriction led to a working sample
  consisting of 6446 CDF sources.

\item Using the redshift information available, we restricted our
  galaxy sample to include only sources with \hbox{$z=0.05-1.2$}.  The
  upper redshift limit corresponds to the maximum distances to which
  we could obtain a complete sample of optically luminous ETGs that
  were relatively bright ($z_{850} < 23$) and contain useful
  morphological information from {\it HST} imaging (see also, e.g.,
  \citealt{Has07}).  Furthermore, this redshift upper limit for our
  survey allows us to detect the majority of \xray\ luminous AGNs with
  $L_{\rm 2-8~keV} \ge 10^{42}$~\ergs\ located in the $\approx$2~Ms
  CDF-N and $\approx$4~Ms CDF-S surveys.  This therefore defines the
  redshift baseline over which we can reliably measure hot gas
  emission through X-ray stacking without significant impact from
  undetected AGNs (see $\S$~4).  As we will discuss below, when
  performing X-ray stacking analyses, we further exclude galaxies with
  $z\leq0.6$ in the more shallow $\approx$250~ks E-CDF-S based on the
  same logic.  For the moment, however, our galaxy sample includes
  E-CDF-S sources with $z \approx$~0.6--1.2, since we will later use
  these galaxies to constrain the radio AGN duty cycle in the ETG
  populations (see $\S$~\ref{sec:energy}).  The imposed redshift
  limits led to the inclusion of 5734 galaxies with $z=0.05-1.2$.

\item Since we are ultimately interested in measuring the hot gas
  X-ray emission from massive ETGs, we required that
  the galaxies that make up our sample have rest-frame $B$-band
  luminosities in the range of $L_B =$~3--30~$\times 10^{10} L_{\rm
    B,\odot}$.  As noted by O'Sullivan et~al.\ (2001; see also
  \citealt{Ellis06} and \citealt{Boroson11}), such optically luminous
 ETGs in the local Universe have relatively massive
  dark matter halos, and are therefore observed to have
  \hbox{0.5--2~keV} emission dominated by hot interstellar gas ($kT
  \simgt$~0.3--1~keV) with minimal contributions from other unrelated
  X-ray emitting sources (e.g., low-mass X-ray binaries; see Fig.~3b).
  This further restriction on including only optically luminous
  galaxies led to 2431 galaxies.

\item To identify passive ETGs in our sample, we made use of the
multiwavelength photometry and redshift information discussed above to measure
rest-frame $U-V$ colours, and we further used $HST$ imaging to provide
morphological information about our galaxies.  As noted by Bell et~al.\ (2004),
the rest-frame $U-V$ colour straddles the 4000~\AA\ break and provides a
sensitive indication of mean stellar age.  For our sample, we first required that
all galaxies have rest-frame $U-V$ colours redder than
\begin{equation}
(U-V)_{\rm rest}=1.15-0.31z-0.08(M_{V}+22.4),
\end{equation}
where $M_{V}$ is the absolute $V$-band magnitude.  Equation~1
(established to be valid out to $z \sim 1$) is based on Bell
et~al.\ (2004; see $\S$~5); however we have used a different constant
term based on our analysis in Fig. \ref{fig:colour_plot}a where we
determine the red/blue galaxy bimodal division by fitting a double
gaussian to the distribution of $U-V+0.31z+0.08 M_{V}$ for our sample
of galaxies after imposing the selection criteria (i) to (iii). In
this exercise, we applied the redshift dependency from \cite{Bell04}
but shifted the constant (by $\sim$$-0.13$) to fit to our sample,
which is consistent with a typical colour scatter of $< 0.2$ mag for
the red sequence colour-magnitude relation (see $\S$~4 in Bell et
al.\ 2004).  We classified galaxies lying below this divide as `blue
cloud' galaxies and those above as `red sequence' galaxies.  We
(A.L.R.D.\ and B.D.L.) then visually inspected all red-sequence
galaxies using grayscale {\it HST} images from the $z_{850}$ band, and
for the subset of sources located in the GOODS-N and GOODS-S
footprints, we also inspected {\it HST} false-colour images based on
$B_{435}$, $V_{606}$, and $z_{850}$ observations.  We strictly
required the galaxies to have bulge dominant optical morphologies for
ETG catalogue inclusion, and we rejected ETG candidates that appeared
to be possible edge-on spirals, which may simply be reddened by
intrinsic galactic dust. Furthermore, we removed five sources which
were very near the edges of the {\it HST} images, where morphological
classification was not possible. Applying these morphological criteria
led to our {\it final sample of 393 optically luminous ETGs}. The
basic properties of our parent sample are shown in Table~\ref{tab:mastertab}.

\end{enumerate}
\begin{table*}
\begin{center}{
\caption{Master Catalogue.}
\begin{tabular}{lllrrrrrlllll}
\hline\hline
RA & Dec & z & spec/phot? & z$_{850}$ & M$_{U}$ & M$_{B}$ & M$_{V}$ & L$_{B}$ & M$_{*}$ & X-ray? & 1.4GHz?  & 24$\mu$m?  \\
(J2000) & (J2000) &  &  &  & & & &  log(L$_{B,\odot}$) & log(M$_{\odot}$) & & & \\ 
(1) & (2) & (3) & (4) & (5) & (6) & (7) & (8) & (9) & (10) & (11) & (12) & (13)  \\
\hline\hline
52.8483000  &  -27.9371400 &  0.816 & p &  21.52 & -21.79 & -22.06 & -22.83 & 11.01 & 11.16 & 0 & 0 & 0 \\
52.8506205  &  -27.9442900 &  1.056 & p &  21.79 & -21.41 & -21.78 & -22.36 & 10.91 & 11.00 & 0 & 0 & 0 \\
52.8527205  &  -27.7069500 &  0.526 & p &  20.62 & -20.49 & -20.83 & -21.65 & 10.52 & 10.84 & 0 & 0 & 0 \\
52.8637605  &  -27.6886300 &  0.908 & p &  21.32 & -21.22 & -21.79 & -22.53 & 10.91 & 11.29 & 0 & 0 & 0 \\
52.8672000  &  -28.0023100 &  0.727 & p &  21.78 & -20.64 & -20.96 & -21.55 & 10.57 & 10.73 & 0 & 0 & 0 \\
52.8714105  &  -28.0047900 &  0.727 & p &  20.82 & -21.33 & -21.56 & -22.24 & 10.81 & 11.04 & 0 & 0 & 0 \\
52.8717405  &  -27.9800800 &  0.771 & p &  21.81 & -21.25 & -21.55 & -22.15 & 10.81 & 10.88 & 0 & 0 & 0 \\
52.8731805  &  -28.0159400 &  0.685 & p &  20.95 & -20.95 & -21.33 & -22.23 & 10.72 & 11.10 & 0 & 1 & 0 \\
52.8741105  &  -28.0181600 &  0.727 & p &  21.64 & -20.81 & -20.99 & -21.67 & 10.59 & 10.80 & 0 & 0 & 0 \\
52.8809595  &  -27.7222000 &  1.005 & p &  22.21 & -21.27 & -21.40 & -21.98 & 10.75 & 11.02 & 0 & 0 & 0 \\
\hline
\end{tabular}

\label{tab:mastertab}
}
\end{center} 
\begin{flushleft}
{\sc Notes}: Columns~(1)--(2): Optical J2000 coordinates. Column~(3):
Source redshift. Column~(4): s=spectroscopic redshift, p=photometric
redshift. Column (5): z$_{850}$ magnitude.  Column~(6): U-band
magnitude. Column~(7): B-band magnitude. Column~(8): V-band
magnitude. Column~(9): Logarithmic $B$-band optical luminosity
(log($L_{B,\odot}$)). Column~(10): Logarithmic stellar mass derived from K-band
magnitude (log(M$_{\odot}$)). Column~(11): Indicates whether the source was
X-ray detected or not (0=not detected, 1=detected). Column~(12): Indicates
whether the source was 1.4GHz radio detected or not (0=not detected,
1=detected). Column~(13): Indicates whether the source was 24$\mu$m
detected or not (0=not detected, 1=detected). Table~\ref{tab:mastertab} is
presented in its entirety (393 sources) in the electronic version of
the journal. Only a portion (first 10 sources) is shown here for guidance.
\end{flushleft}
\end{table*}


We note that out of the 393 optically luminous ETGs that make up our
sample, 190 of these galaxies lie in the CDF-S or CDF-N at
$z=0.05$--1.2, or in the E-CDF-S at $z=0.05$--0.6, which could
potentially be used in X-ray stacking. The remaining 203 sources lie
in the E-CDF-S at $z=0.6$--1.2. Of the 393 galaxies in our sample, 163
have spectroscopic redshifts, and the remaining 230 sources have
high-quality photometric redshifts from the Rafferty et~al.\ (2011)
catalogue. Furthermore, 128 out of 190 sources potentially to be used
in X-ray stacking have spectroscopic redshifts. 

 Using both the photometric and spectroscopic redshifts we carry
  out a basic test of the environment of our sources by searching for
  neighbouring galaxies with an angular separation of $<$500kpc from
  each of our 393 ETGs and within a redshift difference of 0.09 and
  0.046 in the CDF-N and E-CDF-S respectively (the dispersion in the
  photometric redshifts; see footnote 3). We first apply the cut in
  $z_{850}$ optical magnitude of $z_{850}<23$ in order to ensure we
  are only using high quality photometric redshifts and apply a
  further cut in absolute magnitude of M$_{z850}<-21$. This results in
  a total galaxy sample of $\sim3500$ galaxies in the CDF-N and
  $\sim1000$ in the E-CDF-S. We find a median of $10\pm1$ and $4\pm2$
  companions per ETG in the CDF-N and E-CDF-S respectively.  We then
  check the number of neighbours we find for a random galaxy by
  searching within the comparison galaxy catalogues, and find a
  slightly lower median number of companions within 500kpc of $7\pm2$
  and $3\pm1$ for the CDF-N and E-CDF-S respectively.  Therefore, we
  find tentative evidence that the massive ETGs are in richer than
  average environments (likely small groups). Since we are using the
  photometric redshifts there is quite a large uncertainty, however,
  when using only the spectroscopic redshifts (giving us a much
  smaller and likely imcomplete sample) we do still find evidence for
  clustered environments. This is not unexpected since we are
  selecting massive ETGs.

\begin{figure}
\centerline{
\psfig{figure=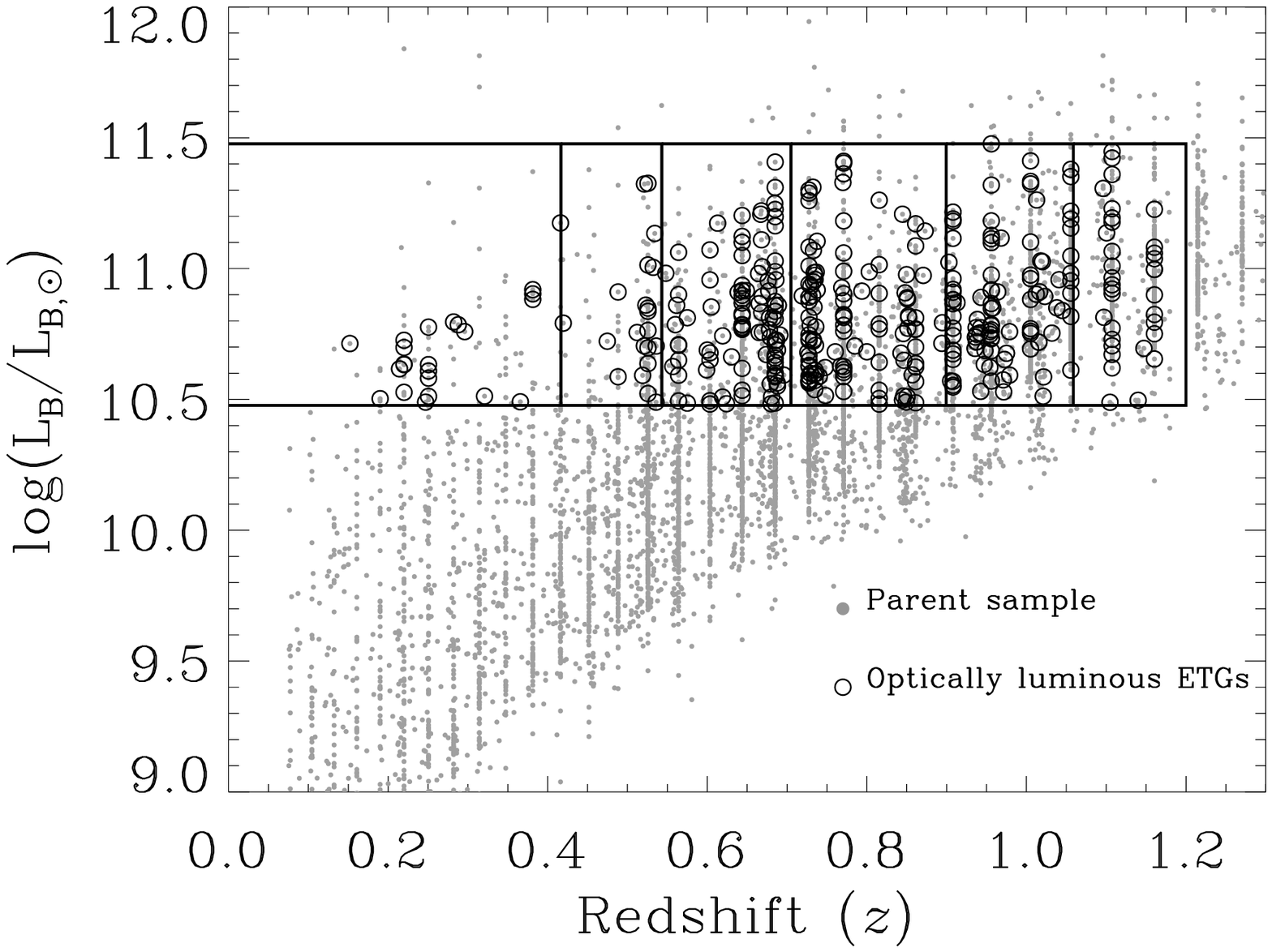,width=3.7in,height=2.7in}}
\caption{The optical $B$-band luminosity of the ETG sample versus the
  redshift.  The entire sample of 5734 sources after imposing the selection
  criteria (i) to (iii) in $\S$ \ref{sample} is represented by filled
  grey circles with the final ETG sample of 393 galaxies as open
  circles.  The boxed areas show the bins from which sources were
  selected for X-ray stacking analysis, described in $\S$
  \ref{sec:Xstack_tech}. The bins were selected to be evenly separated
  in co-moving volume. }
\label{fig:blum}
\end{figure} 

\begin{figure*}
\centerline{\psfig{figure=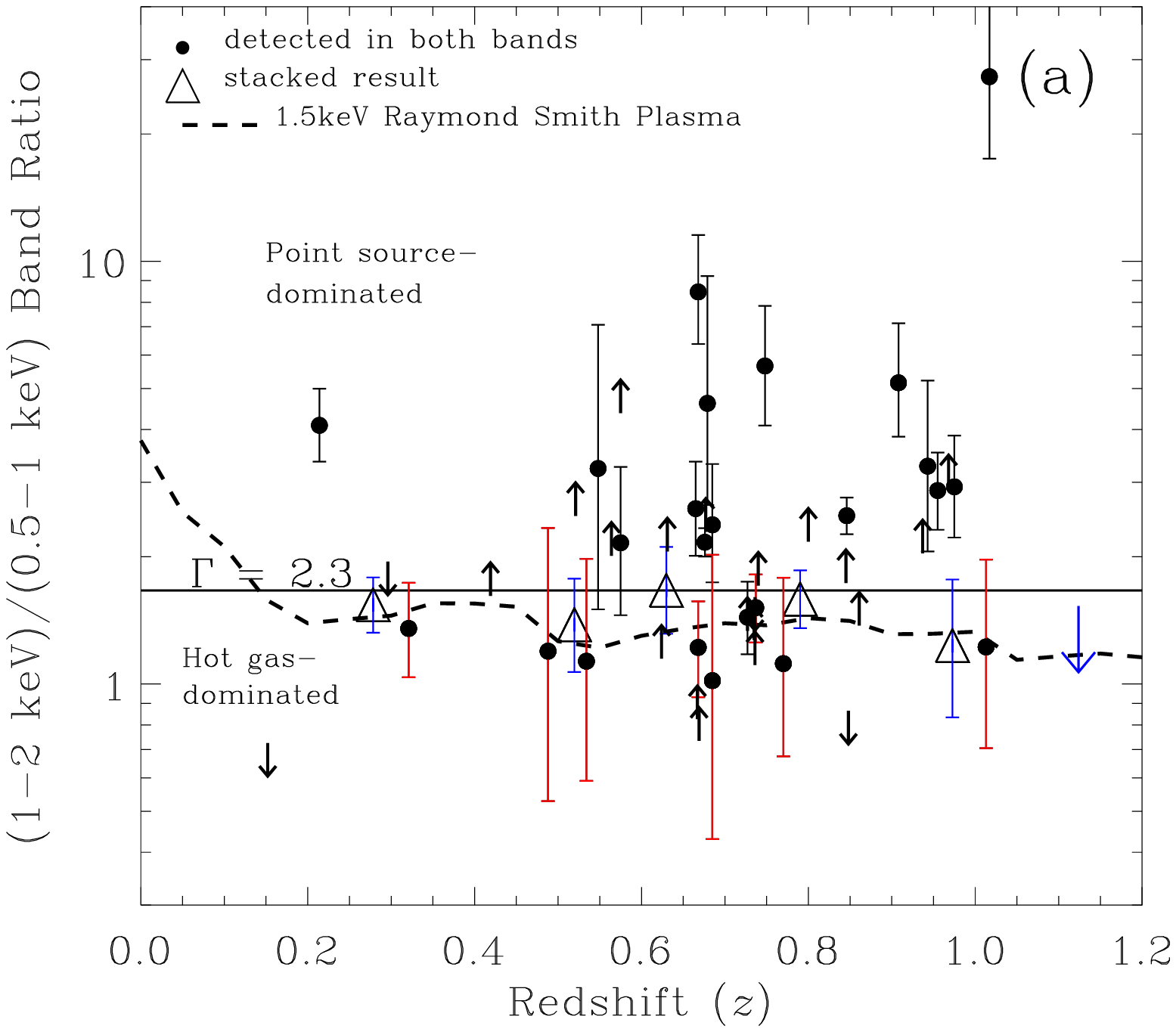,width=3.7in,height=2.7in}
\psfig{figure=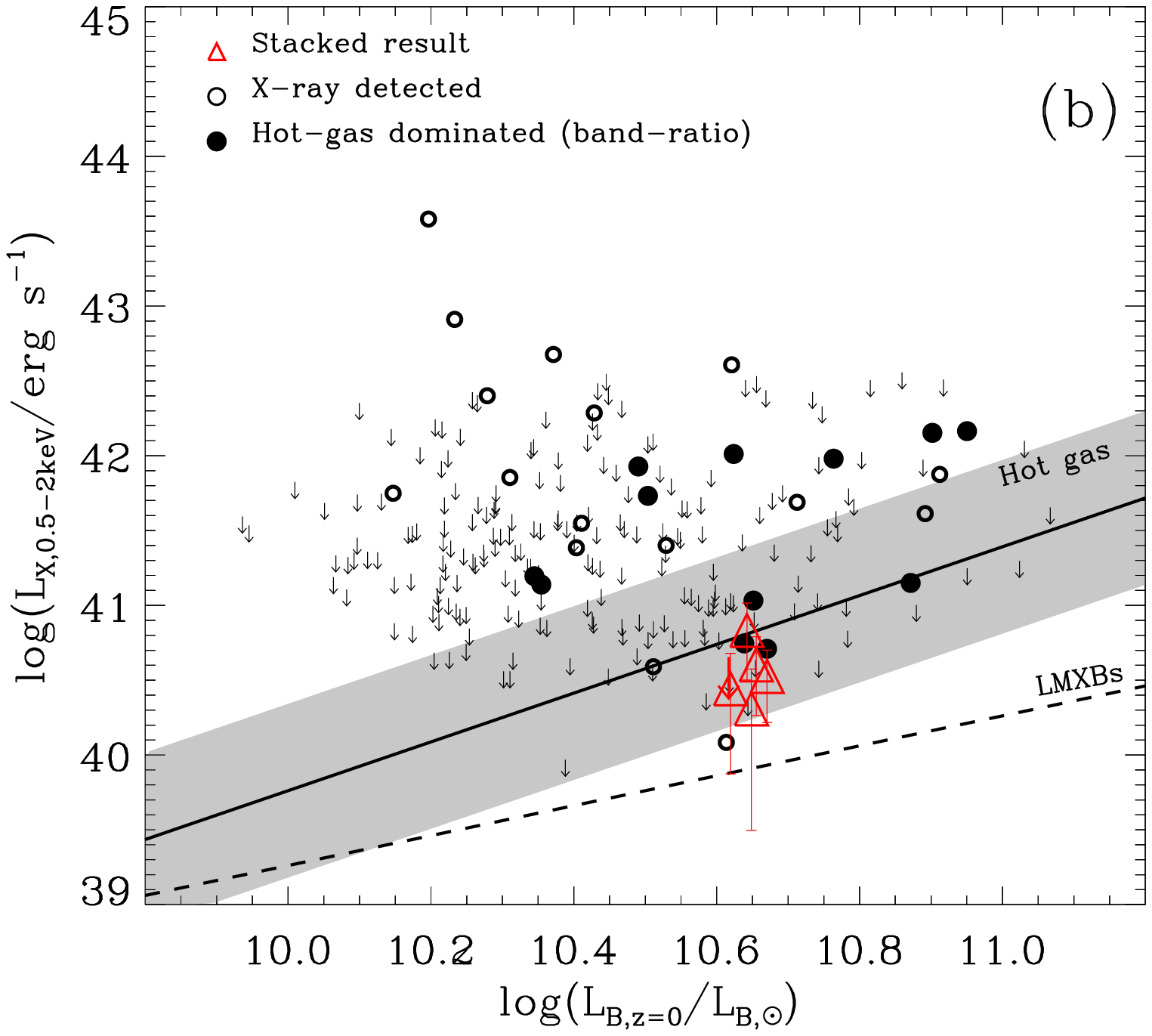,width=3.7in,height=2.7in}}
\caption{(a): The ratio of count-rates in the SB2 and SB1 X-ray bands
  against redshift.  The solid line at $\Gamma_{\rm eff} = 2.3$ shows
  our adopted divide between hot-gas dominated sources and possible
  low mass X-ray binary (LMXB) and/or AGN dominated sources. The SED
  for a 1.5~keV Raymond-Smith plasma (\citealt{RS77}) is shown as a
  dashed curve. Our stacked results (open triangles) are consistent
  with this SED. The errors on the band ratio were determined
  following the `numerical method' described in $\S$~1.7.3 of Lyons
  (1991). Band ratios are corrected for differential vignetting
  between the different bands using the appropriate exposure maps.
  (b): The rest-frame k-corrected SB (0.5--2~keV) X-ray luminosity
  (derived from the 0.5--1~keV flux to minimise contribution from
  LMXBs, and converted to 0.5--2keV luminosity using a 1.5 keV
  Raymond-Smith plasma SED) versus the faded, rest-frame $B$-band
  luminosity (where the evolution of $L_B$ is removed, using Faber
  \etal\ (2007) to parameterise the effect: $L_{\rm B,z=0} = L_{\rm B}
  \times 10^{-0.4 \times 1.23 \times z}$). The solid line and the
  1$\sigma$ dispersion (shaded) show the contribution to X-ray
  emission from hot gas (\citealt{OSullivan01}; converted from
  bolometric to soft-band X-ray luminosity). The dashed line
  represents the expected contribution to X-ray emission from LMXBs
  (\citealt{OSullivan01}); arrows represent the 3$\sigma$ upper limit
  X-ray luminosities. Those sources identified as hot-gas dominated
  from their band ratios are represented here as filled circles, and
  they lie well within 3$\sigma$ of the hot gas relation from
  \citet{OSullivan01}. The results from the stacking procedure in six
  redshift bins are shown as open triangles and these results are
  based on the SB1 (0.5--1~keV) emission, convolved with the SED of a
  1.5~keV Raymond-Smith plasma (\citealt{RS77}) to determine the
  corresponding SB (0.5--2~keV) emission.}
\label{fig:photon-index}
\end{figure*}

\begin{table*}
\begin{center}{
\caption{Hot gas dominated X-ray detected galaxies.}

\begin{tabular}{lrlrrrrlrrl}
\hline\hline
RA & Dec & z & SB1 counts & SB2 counts & SB2/SB1 & L$_{X,SB}$ & $L_{1.4GHz}$ & log(L$_{B,z=0}$) & stack(y/n) \\
(J2000) & (J2000) &  & (0.5--1keV) & (1--2keV) & & $10^{41}$erg s$^{-1}$ & $10^{22}\mu$Jy  & log(L$_{B,\odot}$) & \\ 
(1) & (2) & (3) & (4) & (5) & (6) & (7) & (8) & (9) & (10) \\
\hline\hline
03:32:09.706 &  $-$27:42:48.110 & 0.727 & 74.36$^{+12.16}_{-10.98}$ & 125.44$^{+15.84}_{-14.67}$ & 1.44$^{+0.31}_{-0.26}$ & 185.79$^{+30.38}_{-27.43}$&  56.54$\pm$2.79 & 10.90 & N \\
03:32:28.734 &  $-$27:46:20.298 & 0.737 & 61.89$^{+9.56}_{-8.37}$    &  95.67$^{+11.59}_{-10.41}$  & 1.52$^{+0.30}_{-0.26}$ & 10.17$^{+1.57}_{-1.38}$     &  59.63$\pm$2.81 & 10.62 & Y \\
03:32:34.342 &  $-$27:43:50.092 & 0.668 & 40.22$^{+8.36}_{-7.14}$    &  52.90$^{+9.85}_{-8.64}$     & 1.22$^{+0.35}_{-0.29}$ & 5.19$^{+1.08}_{-0.92}$      & $<$5.39 & 10.50 & Y \\
03:32:38.786 &  $-$27:44:48.923 & 0.736 &  8.96$^{+5.14}_{-3.85}$     &  $<$15.31                           & $<$1.61                        & 1.51$^{+0.87}_{-0.65}$      &  13.10$\pm$1.42 & 10.34 & Y \\
03:32:41.406 &  $-$27:47:17.185 & 0.685 &  7.97$^{+4.82}_{-3.51}$     &   8.41$^{+4.99}_{-3.69}$      & 1.02$^{+1.00}_{-0.59}$ & 1.06$^{+0.64}_{-0.47}$      & $<$5.72 & 10.65 & Y \\
03:32:44.088 &  $-$27:45:41.461 & 0.488 &  8.70$^{+5.15}_{-3.86}$     &  11.10$^{+5.88}_{-4.62}$     & 1.20$^{+1.14}_{-0.67}$ & 0.49$^{+0.29}_{-0.22}$      & $<$2.56 & 10.67 & Y \\
03:32:46.536 &  $-$27:57:13.104 & 0.770 & 25.31$^{+9.69}_{-8.50}$    &  37.30$^{+11.75}_{-10.58}$  & 1.12$^{+0.67}_{-0.44}$ & 68.68$^{+26.29}_{-23.08}$ & $<$7.55 & 10.95 & N \\
03:32:46.949 &  $-$27:39:02.916 & 0.152 & 24.77$^{+8.11}_{-6.90}$    &  $<$25.36                           & $<$0.73                        & 1.32$^{+0.43}_{-0.37}$     &  0.65$\pm$0.04 & 10.64 & Y \\
03:32:52.066 &  $-$27:44:25.044 & 0.534 & 17.67$^{+7.97}_{-6.76}$    &  23.03$^{+9.58}_{-8.39}$     & 1.13$^{+0.84}_{-0.54}$ & 17.88$^{+8.07}_{-6.84}$   &  15.66$\pm1.29$ & 10.87 & Y  \\
12:36:39.760 &   62:15:47.832  & 0.848 & 14.42$^{+5.37}_{-4.21}$    &  $<$11.56                           & $<$0.86                       & 8.41$^{+3.13}_{-2.45}$     & $<$6.34 & 10.49 & Y  \\
12:36:44.414 &   62:11:33.347  & 1.013 & 12.11$^{+5.16}_{-3.89}$    &  14.00$^{+5.37}_{-4.21}$     & 1.22$^{+0.75}_{-0.52}$ & 9.27$^{+3.95}_{-2.98}$     &  876.37$\pm28.74$ & 10.76 & Y  \\
12:36:52.895 &   62:14:44.152  & 0.321 & 35.05$^{+7.37}_{-6.21}$    &  44.53$^{+8.11}_{-6.95}$     & 1.35$^{+0.38}_{-0.32}$ & 1.34$^{+0.28}_{-0.24}$     &  6.41$\pm0.31$ & 10.35 & Y \\

\hline
\end{tabular}

\label{tab:XRayMatches}
}
\end{center} 
\begin{flushleft}
{\sc Notes}: Columns~(1)--(2): Optical J2000 coordinates. Column~(3): Source
redshift. Column~(4): (0.5--1~keV) net counts. Column (5): (1--2~keV) net
counts. Column (6): (1--2~keV)/(0.5--1~keV) count-rate ratio (SB2/SB1).
Column~(7): Rest-frame \hbox{0.5--2~keV} luminosity (ergs s$^{-1}$) derived
from SB1 counts and Raymond-Smith plasma SED. Column~(8): Radio luminosity
($L_{\rm 1.4~GHz}$). Column~(9): Logarithm of the $B$-band optical luminosity
($\log L_{B,\odot}$), Column~(10): Indicates whether the source was included in
our stacking analyses (Y/N).
\end{flushleft}
\end{table*}

%
\section{Multiwavelength Characterisations of ETGs Using Ancillary Data}
\label{obs}

In this section, we make use of the extensive multiwavelength data
available in the CDFs to identify both passive and non-passive (e.g.,
star-forming and AGN) ETGs.  In the next section ($\S$~4), we will
perform \xray\ stacking analyses of the passive ETG population to
measure directly the evolution of the mean hot gas emission.  In the
analyses below, we match our 393 optically luminous ETG optical source
positions to those provided in multiwavelength catalogues using
closest-counterpart matching, which is a reasonable method provided
that the optimum matching radius is carefully selected.  We selected
the optimum matching radius for each multiwavelength catalogue by
first performing matching using a 30\arcsec\ matching radius and then
observing the distribution of closest-counterpart matching offsets.
For all catalogues discussed below (i.e. optical-x-ray, optical-radio,
optical-infrared matching), we found the distribution of offsets to
peak close to $\approx$0\arcsec, reach a minimum at $\approx$1\farcs5,
and subsequently rise toward larger offsets due to spurious matches.
A matching radius of 1\farcs5 was therefore adopted as the optimum
matching radius for all but the radio catalogues, for which the
positional errors are very small, therefore 1\arcsec was more
appropriate. Matches were visually inspected to further ensure they
were sensible. The number of spurious matches was determined for each
data set analytically by calculating the ratio between the total area
covered by the parent sample sources, each with 1\farcs5 or 1\arcsec
matching radius ($\pi (1.5)^{2} \times 393\approx2778$sq. arcsec or
$\pi (1.0)^{2} \times 393\approx1234$sq. arcsec respectively) and the
total area of the CDFs (within 6\arcmin\ of each pointing; 0.188
sq. degrees or 2436480 sq. arcsec). This ratio was then multiplied by
the total number of sources in the multiwavelength catalogues that lie
within 6\arcmin\ of one of the Chandra aimpoints.

\begin{figure}
\begin{center}{
\psfig{figure=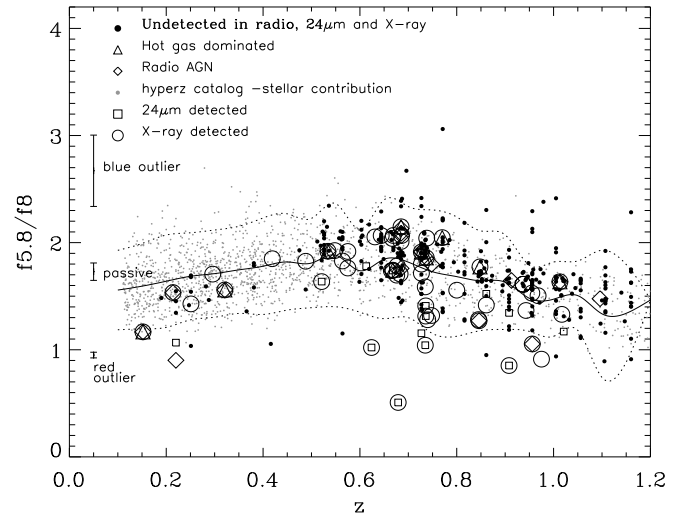,width=3.5in}
\caption{Selecting passive galaxies to use in the X-ray stacking
  analysis based on their IRAC 5.8~$\mu$m/8~$\mu$m colours. The solid
  curve shows the median 5.8~$\mu$m/8~$\mu$m colours of the 5000 model
  passive galaxy SEDs generated by {\it hyperz} with the 2$\sigma$
  dispersion shown as dotted curves. In this figure, the most active
  galaxies (e.g. star-forming galaxies) have redder colours than those
  expected for passive ETGs. The error bars shown at $z=0.05$ are
  representative error bars, plotted at the median 5.8$\mu$m/8$\mu$m
  colour for the following three cases: sources lying within
  $\pm2\sigma$ of the passive line; outliers with lower signal to
  noise with bluer colours ($>2\sigma$ above the passive line); redder
  AGN and active star-forming galaxies ($>2\sigma$ below the passive
  line). In our stacking analysis in $\S$~\ref{sec:Xstack_tech} we
  test the effect of including/not including the sources lying in
  regions which are $>2\sigma$ away from the passive line but which
  have not otherwise been classified as active, however, there is very
  little difference in the final result aside from signal-to-noise. In
  our final stacking analysis presented in Fig. 7 these sources have
  therefore been included.}
\label{fig:IRAC}
}\end{center}
\end{figure}

%
\subsection{X-ray Properties of ETGs}

The ultradeep \chandra\ data in the CDFs provide a direct means for
classifying \xray\ detected ETGs as either hot gas dominated or likely
AGNs.  We used the published main catalogues for each of the CDFs,
which consist of 503 sources in \hbox{CDF-N} ($\approx$2~Ms;
$\approx$0.12 deg$^{2}$ survey, \citealt{Alexander03A}), 740 sources
in the \hbox{CDF-S} ($\approx$4~Ms; $\approx$0.13 deg$^{2}$ survey,
Xue \etal\ 2011), and 762 sources in the \hbox{E-CDF-S} (four
contiguous $\approx$250~ks \chandra\ observations that flank the
\hbox{CDF-S} proper; $\approx$0.31 deg$^{2}$, \citealt{Lehmer05}).
Using our sample of 393 ETGs, the optical coordinates of the galaxies
were matched to the CDF \xray\ catalogue positions using our adopted
matching radius of 1\farcs5.  When an ETG matched to a source in both
the \hbox{E-CDF-S} and the CDF-S simultaneously (due to overlap
between the E-CDF-S and CDF-S) we chose to use the data for the CDF-S,
since these X-ray data are significantly deeper with smaller
positional errors.  In total, 55 X-ray matches were found once
repetitions had been removed, including 11 in the CDF-N and 44 in the
E-CDF-S region.  The fraction of spurious matches in all the CDFs
together was estimated to be $\approx$2.8\% (or $\approx$2 expected
spurious matches).

In Fig.~\ref{fig:photon-index}a, we show the SB2/SB1 count-rate
ratio versus redshift for the \xray\ detected ETGs in our sample.  The
SB2/SB1 ratio provides an effective discriminator of the
\xray\ spectral shape in the SB, the energy regime where hot gas is
expected to dominate. Typically, z$\approx0$-2 AGNs have $\Gamma_{\rm
  eff} \sim 1.8$-2.3 (e.g., \citealt{Alexander05},
\citealt{Vignali02}, \citealt{Reeves00}). Therefore we took the upper
limit of this range and conservatively classified sources with
SB2/SB1~$\le$~1.7 (corresponding to $\Gamma_{\rm eff}\ge2.3$) as
sources having SB emission dominated by a hot gas component.  Sources
detected only in SB2 (i.e., having only a lower-limit on SB2/SB1),
that had SB2/SB1 limits below our adopted cut were not classed as hot
gas dominated sources.  Sources with SB2/SB1 hardness ratio greater
than this cut (i.e., SB2/SB1~$> 1.7$), have X-ray emission likely
dominated by low mass X-ray binaries (LMXBs) or \xray\ AGNs.  However,
by construction, our choice to study optically luminous ETGs will
inherently minimise contributions from LMXB-dominated systems and
therefore AGNs are expected to dominate the SB2/SB1~$> 1.7$ population
(see below).  Our SB2/SB1 criterion indicated 12 hot-gas dominated
sources and 25 likely AGNs (Fig.~\ref{fig:photon-index}a). The SB2/SB1
ratios imply that a Raymond-Smith plasma (\citealt{RS77}) of
kT$\sim$1.5keV is a good spectral model from which to convert
count-rates to flux. In Table~\ref{tab:XRayMatches}, we tabulate the
properties of these \xray\ detected ETGs.

In Fig.~\ref{fig:photon-index}b, we show the \hbox{0.5--2~keV}
luminosity (hereafter, $L_{\rm X}$) versus $L_{B,0}$ (see $\S$~4 for
details) for the ETGs in our sample.  In order to minimise the
contribution from LMXBs we calculated the rest-frame \hbox{0.5--2~keV}
luminosities $L_{\rm X,SB}$ based on the \hbox{0.5--1~keV} SB1 fluxes
provided in the $\Chandra$ catalogues and convert them to
\hbox{0.5--2~keV} SB fluxes, applying a k-correction:
\begin{equation}
L_{\rm X}=4 \pi d_{\rm L}^{2} f_{\rm X} k \;\;\; ({\rm erg~s^{-1}}),
\label{eq:Lx_sb_eq}
\end{equation}
where $d_L$ is the luminosity distance in cm, $f_{\rm X}$ is the
0.5--2~keV flux in units of \ergcms.  The quantity $k$ is the
redshift-dependent k-correction.  For sources that were characterised
as hot gas dominated we used the observed 0.5--1~keV flux and a
Raymond-Smith plasma SED (with $kT_{\rm X} = 1.5$~keV; \citealt{RS77};
see Fig. \ref{fig:photon-index}a) to compute $L_{\rm X}$.  For sources
that were identified as AGN dominant, we used a power-law SED (with
$\Gamma = 1.8$) and the observed 0.5--2~keV flux to compute $L_{\rm
  X}$.  The solid line and shaded region shows the best-fit local
relation and 1$\sigma$ dispersion for hot gas dominated ETGs, and the
dashed line shows the expected contribution from LMXBs (based on
O'Sullivan \etal\ 2001 and typically a factor of $\sim$10 below the
hot gas contribution).  We note that nearly all ETGs without
\xray\ detections (plotted as upper limits) and the 12 ETGs with
SB2/SB1 band ratios consistent with being hot gas dominated
(highlighted with filled circles) also have $L_{\rm X}/L_B$ values
similar to those observed for local hot gas dominated ETGs.  The
majority of the remaining \xray\ detected sources with SB2/SB1~$> 1.7$
are expected to be AGNs.  As Fig.~\ref{fig:photon-index}b shows,
these sources typically have large values of $L_{\rm X}/L_B$, again
consistent with that expected from AGNs (see \citealt{OSullivan01};
\citealt{Ellis06}). To further check for AGNs in our sample we
cross-matched our optical catalogue with spectroscopic data from
\cite{Szokoly04}, \cite{Mignoli05}, \cite{Ravikumar07},
\cite{Boutsia09} and \cite{Silverman10} using a 1\farcs5 radius in
order to identify any sources with spectral features indicative of
AGN, such as broad emission lines. We identified two potential broad
line AGN in the E-CDF-S, both of which were X-ray detected and had
already been flagged as likely AGN using our band ratio analysis
(Fig.~\ref{fig:photon-index}).

\begin{figure*}
\centerline{
\psfig{figure=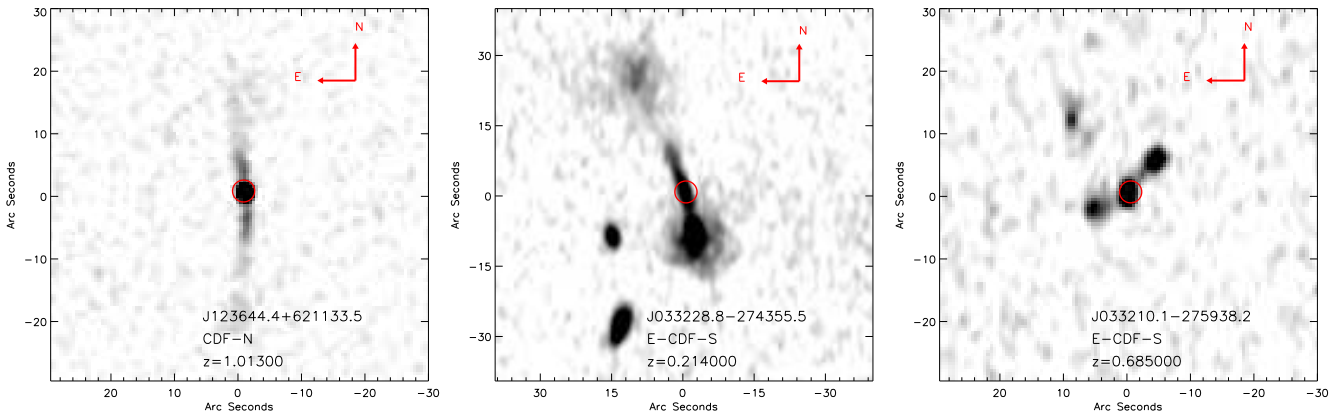,width=1.0\linewidth,angle=90}}
\caption{Radio images of the three extended radio matches discovered by visual
inspection of the radio images. The axes give the distance to the centre of the
image in arcseconds. All of these extended sources have significantly higher
1.4GHz fluxes (1.2--4.8mJy) than the other 16 radio-bright AGN in our ETG
sample.}
\label{fig:radio_sources}
\end{figure*}

%
\subsection{5.8--24~$\mu$m Properties of ETGs}

In order to explore further whether the ETGs contained more subtle
signatures of AGN or star-formation activity than provided by their
\xray\ and optical spectroscopic properties, we utilised
\Spitzer\ photometry over the 5.8--24~$\mu$m range.  We began by using
\Spitzer\ IRAC 5.8~$\mu$m/8~$\mu$m colours.  Since AGN tend to be redder
than galaxies in the mid-infrared, the 5.8~$\mu$m/8~$\mu$m colour can be
used to identify AGNs when the continuum is dominated by a rising
power law component rather than a dropping stellar component (e.g.,
\citealt{Stern05}).  Similarly, the SEDs of powerful star-forming
galaxies, containing a large hot dust component, may exhibit this rise
towards redder wavelengths and may also be identified by their
5.8~$\mu$m/8~$\mu$m colour.  In the \hbox{E-CDF-S}, we take the
photometry in these channels from \cite{Damen11} and in the CDF-N,
we take photometry from \hbox{GOODS-N}, cutting both catalogues at a
signal-to-noise level of
S/N$<5$. \footnote{http://data.spitzer.caltech.edu/popular/goods/}

We matched the positions from our sample of 393 ETGs to the
\Spitzer\ IRAC catalogues using a matching radius of 1\farcs5 and
found 384 matches.  We estimated a spurious matching fraction of
$\approx$3.1\% ($\approx$12 matches).  In Fig.~\ref{fig:IRAC}, we show
the 5.8~$\mu$m/8~$\mu$m colour versus redshift for the 384 ETGs in our
main sample.  To determine the expected 5.8~$\mu$m/8~$\mu$m colours
for passive galaxies, we used the code {\ttfamily hyperz}
\citep{Bolzonella00} with the SED library of \cite{Bruzual93} to
generate 5000 model galaxy SEDs based on a wide range of star
formation histories and redshifts.  For each galaxy, we adopted a
formation redshift randomly selected to lie between the galaxy
redshift (i.e., $t_{\rm age} = 0$) and $z \approx 5$.  In
Fig.~\ref{fig:IRAC}, we plot the running median of the
5.8~$\mu$m/8~$\mu$m colours for the {\ttfamily hyperz} sample, after
imposing the rest-frame $U-V$ color criterion in equation~1 (the solid
line in Fig.~\ref{fig:IRAC}), and calculate the 2$\sigma$ dispersion
either side of the median (the dashed lines on Fig.~\ref{fig:IRAC}).
This curve was calculated by binning the data into bins of $\Delta
z=0.05$ and computing the median and dispersion for each bin.  Sources
with very red IRAC colours lying below the passive line are likely AGN
or star forming galaxies, and it can be seen that X-ray, radio and
24~$\mu$m detected galaxies tend to lie below the solid line. In our
stacking analyses (see $\S$~\ref{sec:Xstack_tech}), we experimented
with removing the sources that lie outside of the 2$\sigma$ dispersion
boundaries of our hyper-z normal galaxy envelope.  However, we found
no difference in the general results since most of the sources
exhibiting non-passive activity (either due to star formation or AGNs)
had already been identified by other indicators, and we therefore
decided to include all of our galaxies in our subsequent stacking
analyses (unless otherwise flagged as non-passive). As an additional
test we experimented with the IRAC colour-colour diagnostic as in
\cite{Stern05}, Fig.~1, however, we find that only three of our
sources lie in their region of active sources, all three of which we
have already flagged as active sources through our other diagnostics.

To identify additional ETGs in our sample that have signatures of star
formation or AGN activity from dust emission, we cross-matched our ETG
sample with \Spitzer\ MIPS 24~$\mu$m catalogues.  The \hbox{E-CDF-S}
was observed with \Spitzer/MIPS as part of the FIDEL legacy program
\footnote{http://irsa.ipac.caltech.edu/data/SPITZER/FIDEL/} (PI: Mark
Dickinson; see $\S$~2.1.2--2.1.3 of Magnelli \etal\ 2009).  We have
used a catalogue of 20329 sources produced by the DAOPHOT package in
IRAF (see $\S$~2.3 of Biggs \etal\ 2011).  The MIPS 24~$\mu$m
sensitivity over the E-CDF-S varies significantly across the
$30\arcmin \times 30\arcmin$ field, with exposure times ranging from
11,000--36,000~s.  We make use of sources having signal-to-noise of at
least 5$\sigma$ ($\approx$30--70~$\mu$Jy limits; see Magnelli
\etal\ 2009).  For the CDF-N, we made use of the publicly available
GOODS \Spitzer\ Legacy survey catalogues of 1198 sources (PI:
M.~Dickinson).  We utilised the 5$\sigma$ sample (flux limits of
70~$\mu$Jy in the E-CDF-S and 30~$\mu$Jy in the CDF-N;
\citealt{Magnelli09}).  Using a 1\farcs5 matching radius we found a
total of 20 matches to the 393 ETGs in our sample; three in the CDF-N
and 17 in the E-CDF-S, with $\sim$1.7 spurious matches expected.
24~$\mu$m provides a robust diagnostic of the presence of cold dust
emission from the circumstellar envelopes of young embedded
UV-luminous stars, characterised by a rising SED through the
mid-infrared \cite{Muzerolle04}.  Such systems are expected to contain
significant \xray\ contributions from populations that are unrelated
to hot gas, and we therefore classify these 20 sources to be
star-formation active systems.

\subsection{Radio Properties of ETGs}

To measure powerful radio emission produced by either radio-loud AGNs
or star-formation activity, we cross-matched the optical coordinates
of the parent sample with 1.4~GHz VLA catalogues in the CDFs (using a
1\arcsec\ matching radius).  We utilised the catalogue from
\cite{Miller08}, but included additional sources at S/N$>$5 (Miller et
al. in preparation) for the \hbox{E-CDF-S} region, which contains 940
radio sources with S/N$>5$ and reaches a 5$\sigma$ limiting flux
density of 30~$\mu$Jy with a synthesised beam of 2.8\arcsec\ $\times$
1.6\arcsec. For the \hbox{CDF-N}, we utilised the catalogue from the
\cite{Morrison10} \hbox{GOODS-N} observations, which provides entries
for 1227 discrete radio sources with S/N$>5$ and 5$\sigma$ flux
density limit of 20~$\mu$Jy at the field centre, with a
1.7\arcsec\ beam.  In total, 24 radio detected counterparts to the 393
ETGs were found (six in the CDF-N and 18 in the E-CDF-S) and 15 of
these radio detected ETGs were also X-ray detected.  The spurious
matching fraction was estimated to be $\approx$0.5\% ($\approx$0.1
matches) and therefore negligible.  Since the radio emission from
radio luminous AGNs can be extended (e.g., \citealt{FR74}), the radio
maps were carefully inspected by eye against the 1\arcsec\ radius
matching circles (overlaid at the locations of the parent sample
positions) to verify the accuracy of the matches and isolate extended
sources.  We identified three bright extended sources that were all
identified using closest-counterpart matching; radio images of these
sources have been provided in Fig.~\ref{fig:radio_sources}. We note
that some of these individual sources have been well studied in the
literature (e.g. J123644.4; \citealt{Richards98}, J033238.8 and
J033210.1; \citealt{Kellerman08}).

We calculated rest-frame 1.4~GHz monochromatic luminosities for all
radio detected sources following,
\begin{equation}
L_{\rm 1.4~GHz}=4\pi d^{2}_{\rm L} f_{\rm 1.4GHz} 10^{-36} (1+z)^{\alpha-1}
{\rm W~Hz^{-1}}, 
\end{equation}
where $f_{\rm 1.4~GHz}$ is the 1.4~GHz flux density ($\mu$Jy) and
$\alpha$ is the radio spectral index for a power-law radio SED (i.e.,
$F_\nu \propto \nu^{-\alpha}$).  We adopted a power-law spectral index
of $\alpha=0.85$ (see \citealt{Richards00} for motivation).  For
normal galaxies without active radio AGNs, radio emission originates
from HII regions and Type~II and Ib supernovae, which produce
synchrotron radiation from relativistic electrons and free-free
emission (\citealt{Condon92}).  In passive ETGs, the contribution from
these processes is unlikely to exceed $10^{18-19} $WHz$^{-1}$
(\citealt{Ledlow97}).  The radio luminosity for all radio detected
sources in our sample was greater than $10^{20}$~W~Hz$^{-1}$, which is
expected given the flux limits of our survey. Therefore, detecting
them at all suggests an excess of non-passive activity from either
star-formation (SFR~$\simgt 0.1$~$M_\odot$~yr$^{-1}$) or AGN activity.

To discriminate between star-formation and AGN activity in the
radio-detected population, we use the well-known strong correlation
between radio and far-infrared emission, which extends to
cosmologically significant redshifts (at least $z \approx 1$;
\citealt{Appleton04}, and z$>2$ using total infrared luminosity;
\citealt{Mao11}).  For all our ETGs that are detected at both
24~$\mu$m and 1.4~GHz we measured the quantity $q_{24} \equiv \log
(f_{24 \mu m}/f_{\rm 1.4GHz})$ \citep{Appleton04} (where $f_{24 \mu
  m}$ and $f_{\rm 1.4GHz}$ are observed fluxes).  Radio-excess AGN can
be identified by comparing their infrared emission to their radio
emission, as their radio emission is significantly brighter than their
infrared emission when compared to star-forming galaxies, which fit
tightly along the far-infrared/radio correlation.  Following Del Moro
et al.  (submitted) demonstrating the typical $q_{24}$ of radio-excess AGN
based on starburst SEDs, we apply a selection of $q_{24} < 0.5$ to be
indicative of radio AGN.  This results in 19 of the 24 radio-detected
galaxies being classified as radio AGN, with the remaining five
radio-detected galaxies being classified as star-forming galaxies (as
indicated in Table \ref{tab:RadioMatches}).  In this exercise 10 ETGs
with 24~$\mu$m detections but not radio detections were excluded from
the final sample. We note that 16 of the 24 radio AGN were also X-ray
detected.  Table~\ref{tab:RadioMatches} shows the matched radio
sources that are classified as radio AGN from the q$_{24}$ analysis,
and which are used to estimate the AGN heating in $\S$~5. We note that
this approach identified all the sources with extended radio emission
in Fig.~\ref{fig:radio_sources} as radio AGN.

In Table~\ref{tab:summary}, we summarise the various source
classifications described in $\S$~3 for clarity. Of the original 393
galaxies in the ETG sample 190 of them can potentially be used in the
X-ray stacking (see $\S$~2). However, through various classification
schemes we find that 32 of these are non-passive (potential X-ray AGN
or star-forming galaxies) and are therefore excluded from the main
stacking sample, thus leaving a sample of 158 passive galaxies which
are suitable for X-ray stacking analyses. Of the 393 ETGs, 24 are
radio detected and 20 of these are likely radio AGN while the other
five have radio emission dominated by star formation. We classify a
further 10 sources as likely star-forming galaxies, which have
detections only in 24~$\mu$m and not radio, and lower limits of
q$_{24}>0.5$.

\begin{table}
\begin{center}{
\caption{Summary of source classifications.}

\begin{tabular}{ll}
\hline\hline
Classification & No. of galaxies \\
\hline\hline
ETG sample & 393 \\
X-ray detected & 55 \\
Passive X-ray detected & 12 \\
Potential LMXB/X-ray AGN & 43 \\
Radio detected & 24 \\
Radio AGN & 19 \\
24$\mu$m detected & 20 \\
Star-forming galaxies  & 15 \\
X-ray stacked galaxies (main) & 158 \\
X-ray stacked galaxies (faded) & 60 \\
\hline
\end{tabular}

\label{tab:summary}
}\end{center} 
\end{table}

\begin{figure}
\centerline{
\psfig{figure=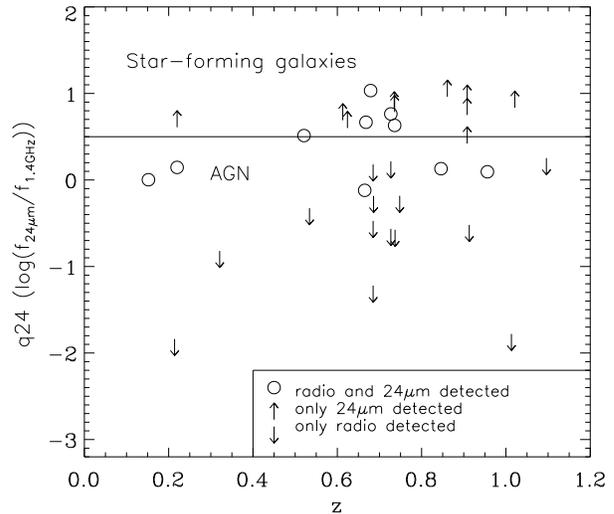,width=3.5in}}
\caption{Ratio of observed flux in 24~$\mu$m and radio (1.4GHz) versus
  redshift. Sources which lie below the solid line (Del Moro et al, in
  prep) are taken to be radio AGN.}
\label{fig:q24}
\end{figure}

\begin{table*}
\begin{center}{
\caption{Radio-bright ETGs.}
\begin{tabular}{lrlrlrrrlll}
\hline\hline
RA & Dec & z & $f_{1.4\rm GHz}$ & log(L$_{1.4GHz}$) & $f_{\rm 24 \mu m}$ & q$_{24}$ & L$_{X,SB}$ & log(L$_{B}$) & Extended & Note \\
(J2000) & (J2000) & & ($\mu$Jy) & ($10^{23}$WHz$^{-1}$) & ($\mu$Jy) & & ($10^{42}$erg s$^{-1}$cm$^{-2}$) & (log(L$_{B,\odot}$)) & (Y/N) & (AGN/SF)\\ 
(1) & (2) & (3) & (4) & (5) & (6) & (7) & (8) & (9) & (10) & (11) \\
\hline\hline  
 03:31:29.563 & -28:00:57.384 & 0.685   &   46.3$\pm$8.8       & 0.9$\pm$0.2     &  $<$70                     &    $<$0.18    & ...   & 10.7  & N & A \\
 03:31:32.210 & -27:43:08.076 & 0.956   &   68.7$\pm$7.3       & 2.9$\pm$0.3     &   85.5$\pm$6.2    &    0.09    & $<$5.1  & 10.9 &  N & A\\ 
 03:31:39.041 & -27:53:00.096 & 0.220   &   61.2$\pm$7.0       & 0.08$\pm$0.01 &   85.5$\pm$5.3    &    0.15    & ...   & 10.6  & N & A \\
 03:31:40.044 & -27:36:47.628 & 0.685   &  208.1$\pm$15.4    & 4.0$\pm$0.3     &   $<$70                    &    $<$-0.47    & ...   & 11.2  & N & A\\
 03:31:45.895 & -27:45:38.772 & 0.727   &   42.7$\pm$6.8       & 0.9$\pm$0.2     &   $<$70                    &    $<$0.22    & ...   & 11.3 & N & A\\
 03:31:57.782 & -27:42:08.676 & 0.665   &   97.2$\pm$6.5       & 1.7$\pm$0.1     &   73.7$\pm$3.1    &  -0.12    & 8.4  & 10.8 & N & A \\
 03:32:09.706 & -27:42:48.110 & 0.727   &  257.2$\pm$12.7    & 5.6$\pm$0.3     &   $<$70                    &   $<$-0.57   & 18.6  & 11.3 & N & A\\
 03:32:10.137 & -27:59:38.220 & 0.685   & 1165.0$\pm$36.0   & 22.2$\pm$0.7   &   $<$70                    &   $<$-1.22   & ...  & 11.2 & Y & A \\
 03:32:19.305 & -27:52:19.330 & 1.096   &   39.1$\pm$6.2       & 2.3$\pm$0.4     &   $<$70                    &    $<$0.25   & ... & 11.3 &  N & A \\
 03:32:28.734 & -27:46:20.298 & 0.737   &  263.3$\pm$12.4    & 6.0$\pm$0.3     &   $<$70                    &   $<$-0.58   & 1.0 &  11.0 & N & A \\
 03:32:28.817 & -27:43:55.646 & 0.214   & 4814.0$\pm$103.0 & 6.3$\pm$0.1     &   $<$70                    &   $<$-1.84   & 0.04  & 10.6 & Y & A \\
 {03:32:38.786} & {-27:44:48.923} & {0.736}   &   {58.0$\pm$6.3}       & {1.3$\pm$0.1}     &   {247.8$\pm$2.4}  &     {0.63}   & {0.2}  & {10.7} & {N} & S \\
 03:32:39.485 & -27:53:01.648 & 0.686   &  107.0$\pm$6.2      & 2.04$\pm$0.12     &   $<$70                    &   $<$-0.18   & $<$0.2 & 11.0 & N & A \\
 03:32:46.949 & -27:39:02.916 & 0.152   &  105.5$\pm$7.0      & 0.060$\pm$0.004   &  106.2$\pm$2.7   &     0.003   & 0.1 &  10.7 &  N & A \\
 {03:32:48.177} & {-27:52:56.608} & { 0.668}   &   { 32.8$\pm$6.2}       & { 0.6$\pm$0.1}     &  { 152.4$\pm$2.3}   &     { 0.67}   & { 4.9}  & { 11.2} & {N} & S \\
 03:32:52.066 & -27:44:25.044 & 0.534   &  148.3$\pm$12.2    & 1.6$\pm$0.1     &  $<$70                     &   $<$-0.326   & 1.79 &  11.1 & N & A\\
 {03:33:05.671} & {-27:52:14.268} & { 0.521 }  &   { 55.7$\pm$6.8    }   & { 0.6$\pm$0.1  }   &  { 181.5$\pm$7.6}   &     { 0.51}   & {$<$2.4}  & { 11.3} & {N} & S \\
 {03:33:15.427 } & {-27:45:24.012 } & { 0.727 }  &   { 63.5$\pm$6.9  }     & { 1.4$\pm$0.2 }    &  { 368.3$\pm$2.7 }  &     { 0.76 }  & ...  & { 10.6} & {N} & S \\
 12:36:01.813 &  62:11:26.659 & 0.913   &   99.2$\pm$5.5        & 3.8$\pm$0.2     &  $<$30                     &    $<$-0.52  &  ... &  10.9 & N & A\\
 {12:36:08.137} &  { 62:10:36.136} & { 0.679 }  &  { 213.1$\pm$7.9  }     & { 4.0$\pm$0.2 }    &  { 2300.0$\pm$12.9} &    { 1.03 } & { 0.2 } &  { 10.7} & {N} & S \\
 12:36:17.098 &  62:10:11.554 & 0.846   &   65.3$\pm$8.3        & 2.1$\pm$0.3     &  88.2$\pm$6.4     &      0.13  & 7.95  & 10.7  & N & A\\
 12:36:22.705 &  62:09:46.313 & 0.748   &   45.7$\pm$5.1        & 1.1$\pm$0.1     &  $<$30                     &    $<$-0.18   &  ...       & 10.6 &  N & A\\
 12:36:44.414 &  62:11:33.347 & 1.013   & 1805.1$\pm$59.2       & 87.6$\pm$2.9   &  $<$30                     &    $<$-1.78   & 0.9  & 11.3 &  Y & A\\
 12:36:52.895 &  62:14:44.152 & 0.321   &  198.3$\pm$9.6        & 0.64$\pm$0.03     &  $<$30                     &    $<$-0.82   & 0.1  & 10.5 & N & A\\
\hline
\end{tabular}

\label{tab:RadioMatches}
}\end{center} 
\begin{flushleft}
{\sc Notes}: Columns (1)--(2): Optical J2000 coordinates. Column (3):
Redshift. Column (4): Radio (1.4GHz) flux density ($\mu$Jy). Column~(5): Radio
(1.4~GHz) luminosity ($\log 10^{23}$W Hz$^{-1}$). Column~(6): 24~$\mu$m flux
density ($\mu Jy$). Column~(7): $q_{24}$ ratio, ($\log f_{24 \mu
m}/f_{1.4GHz}$).  Column (8): 0.5--2~keV flux
($10^{42}$~ergs~s$^{-1}$~cm$^{-2}$) derived from SB1 (0.5--1~keV) counts and
converted using the 1.5~keV Raymond-Smith plasma. Column~(9): $B$-band
luminosity ($\log L_{\rm B,\odot}$).  Column~(10): Indicates whether there is
extended emission (Y=yes, N=no). Column~(11): Note on classification: S = radio
detected ETGs for which the radio emission is likely dominated by star
formation (as implied by their $q_{24}$ value), A = radio emission dominated by
an AGN.
\end{flushleft}
\end{table*}

%
\section{Cosmic History of X-ray Emission from Massive ETGs}
%

Approximately 90\% of the ETGs in our passive sample are undetected in
the X-ray (338 galaxies). Therefore to measure the hot gas emission
from the whole population it is necessary to implement X-ray stacking
techniques. This investigation focuses on measuring the cooling of the
hot gas in ETGs, which dominates emission at soft X-ray energies
(0.5--2~keV), as opposed to LMXBs, which dominate emission in the hard
band (2--8~keV). Stacking analyses were therefore carried out in the
soft bands (i.e., SB1 and SB2), which we expect to be dominated by
hot-gas emission and to have minimal contributions from LMXBs. In
Fig.~\ref{fig:blum}, we plot six redshift intervals of galaxies with
$L_B = (3-30) \times 10^{10} L_{B, \odot}$ where we performed stacking
analyses for each subsample (solid boundaries).  The redshift
divisions were chosen to encompass roughly equal intervals of comoving
volume, and the larger redshift interval spacing beyond $z \approx
0.6$ is the result of excluding from our stacking analyses sources
that were within the $\approx$250~ks exposure of the E-CDF-S (see
$\S$2 for details).

We note that previous studies (e.g., Bell \etal\ 2004; Faber
\etal\ 2007) have shown that, from $z=1-0$, the $B$-band luminosity of
typical massive ETGs fades by $\sim$1~mag.  To estimate the mean X-ray
luminosity evolution for an ETG population with similar $z = 0$
$B$-band luminosities ($L_{B,z=0}$), we thus constructed six ``faded''
redshift-divided subsamples of ETGs with $L_{B,z=0} = (3-30) \times
10^{10} L_{B, \odot}$.  We calculated $L_{B,z=0}$ following the Faber
\etal\ (2007) prescription: $L_{B,z=0} = L_B \times
10^{-0.4\times1.23\times z}$. With these faded luminosities only 60 of
the 158 stacking sources lay within the allowed range of optical
luminosities. In total, we stacked 12 subsamples of ETGs (six main and
six faded) with both the main and faded samples having the same
divides in redshift but with 158 sources in the total main sample and
only 60 in the total faded sample.

\subsection{X-ray Stacking Technique}
\label{sec:Xstack_tech}

Our stacking procedure, summarized below, makes use
of images, background maps, and exposure maps that were constructed by
Alexander \etal\ (2003) for the $\approx$2~Ms CDF-N, Lehmer \etal\ (2005) for
the $\approx$250~ks E-CDF-S, and Xue \etal\ (2011) for the $\approx$4~Ms CDF-S.

We chose to use circular apertures of constant radii to extract
on-source counts.  We chose to extract X-ray counts (source plus
background) from a 1\farcs5 radius circular aperture centered on the
locations of sources that were within 6\arcmin\ of any of the six
\chandra\ aimpoints (the $\approx$2~Ms CDF-N, the $\approx$4~Ms CDF-S,
and the four $\approx$250~ks pointings in the E-CDF-S).  These choices
of source inclusion radius and extraction aperture radius were
previously found to optimise the stacked signal (see, e.g., Lehmer
\etal\ 2005b, 2007) and are therefore implemented here.  For each
source, we used our source extraction aperture to extract source plus
background counts $s_i$ from images and exposure times $t_i$ from
exposure maps.  For each stacked sample, total source plus background
counts were computed as $S = \sum_i s_i$ and exposure times were
computed as $T = 0.03 \sum_i t_i$ (the 0.03 factor comes from the fact
that $t_i$ is the sum of exposure map values over $\approx$30 pixels).

Background and exposure maps were then used to measure the background
counts and exposures for each source.  For this exercise, we used a
15\arcsec\ radius circular aperture centred on the location of each
source to extract local background counts $b_{i, \rm local}$ and
exposure times $t_{i, \rm local}$.  The on-source background counts
$b_i$, were estimated following $b_i = b_{i, \rm local} \times
t_i/t_{i,\rm local}$.  Total stacked background counts were then
obtained through the summation $B = \sum_i b_i$.

For each stacked sample, any galaxy that was classified as a normal
ETG (see $\S$3) was stacked.  The stacking procedure was carried out
with three different samples: (a) a sample including all radio AGN,
passive X-ray detected sources (10 galaxies) and passive X-ray
undetected sources; (b) a sample including only X-ray undetected
galaxies and (c) a sample including passive X-ray undetected galaxies
and radio AGN but excluding passive X-ray detected galaxies.  However,
we found that the inclusion of both X-ray detected normal galaxies and
radio AGN in the stacking did not significantly change our results,
implying that most X-ray luminous AGN had been successfully excluded
from the sample via direct X-ray detection and
classification. Therefore all radio AGN and X-ray detected normal
galaxies (thus all passive galaxies) were included in all of the
stacks resulting in a final sample of 158 and 60 passive ETGs to be
stacked in the main and faded samples, respectively.

For each stacked sample, we measured the signal-to-noise ratio ($S/N =
(S-B)/\sqrt{B}$).  For a significant detection, we required that $S/N
> 3$, and for such stacks, we measured net counts as $N = S-B$.  When
a stacked sample was not detected, we placed 3$\sigma$ upper limits on
the net counts i.e. $N < 3\sqrt{\alpha_{S}^2+\alpha_{B}^2}$, where
$\alpha_{S}$ and $\alpha_{B}$ are the bootstrapped errors on the total
and background counts respectively.  The error on the net counts was
determined by applying a bootstrapping method.  For each stacked
sample containing $n$ galaxies, we randomly drew $n$ sources from the
sample (allowing for multiple draws of the same source) and restacked
the scrambled sample to measure net counts.  This exercise was
performed 1000 times for each stacked sample, thus giving a sense of
the variance of the population.  The count-rates $\Phi$ for each
stacked sample were determined as $\Phi = \xi N/T$, where $\xi$ is a
mean aperture correction.  Since many of the sources with relatively
large PSFs (at $\approx$3--6\arcmin\ off axis) had aperture radii that
did not encompass the whole PSF, it was necessary to factor in a
correction ($\xi_i$) for each of the stacked sources.  The average
correction factor used was computed as $\xi \equiv \sum_i \xi_i
t_i/\sum_i t_i$ where $\xi_i$ and $t_i$ are the correction factors and
exposure times measured for each individual source.  The stacked
count-rates $\Phi$ were then converted to fluxes using the SED for a
1.5 keV Raymond-Smith plasma (\citealt{RS77}; see Fig.~3a for
motivation and see Table~\ref{tab:SB1stack} for kT$_{X}$ values used in each stack
sample).  Errors on the count-rate to flux conversion were calculated
by propagating errors on the mean SB2/SB1 ratio, which can be used as
a proxy for temperature in our Raymond-Smith SED. The errors on the
luminosity were determined by propagating the bootstrapped errors on
the source counts and the errors on the countrate-to-flux conversion
factor based on the errors in the SED temperature described
above. When calculating luminosities, the luminosity distance is
calculated using the mean redshift in each bin (since as Fig.~
\ref{fig:blum} demonstrates, the redshifts are quite evenly
distributed in each stacking bin).

\begin{table*}
\begin{center}{
\caption{X-ray stacking properties.}
\begin{tabular}{llllllllllllll}
\hline\hline
 &  & &  & \multicolumn{2}{|c|}{Exposure time} & &  & &  &   & &  \\  \cline{5-6} 
&  & \multicolumn{2}{|c|}{Net counts (S-B)} & SB1  & SB2 & \multicolumn{2}{|c|}{S/N}  & f$_{SB1}$  & kT$_{X}$ & L$_{X,SB}$   & L$_{B,mean}$ &  L$_{X,SB}$/L$_{B}$ \\ \cline{3-4}  \cline{7-8}
z$_{mean}$ & N$_{tot}$ & SB1 & SB2 & (Ms) & (Ms) & SB1  & SB2 & log(erg s$^{-1}$cm$^{-2}$) & (keV) & log(erg s$^{-1}$) & log(L$_{B,\odot}$) & log(erg s$^{-1}$L$_{B,\odot}^{-1}$) \\ 
(1) & (2) & (3) & (4) & (5) & (6) & (7) & (8) & (9) & (10) & (11) & (12) & (13) \\
\hline
& & & & & \multicolumn{3}{|c|}{General Sample} & & & & &  \\
\hline
 0.278 & 19 &  80.7$\pm$42.2 & 125.3$\pm$66.3 & 12.2  &  12.3  &  29.3 &  36.4   &  -16.1  & 1.3$^{+0.2}_{-0.1}$ & 40.7$\pm0.2$   &  10.7 & 29.9$\pm0.2$      \\
 0.519 & 19 &  38.8$\pm$20.3 & 55.0$\pm$28.1  & 25.7  &  26.3  &  9.9  &  11.1   &  -16.8  & 1.3$^{+0.6}_{-0.1}$ & 40.6$\pm0.2$   &  10.8 & 29.8$\pm0.2$      \\
 0.630 & 33 &  58.2$\pm$40.5 & 97.3$\pm$50.2  & 85.2  &  85.5  &  7.7  &  10.2   &  -17.0  & 1.6$^{+0.5}_{-0.4}$ & 40.5$\pm0.3$   &  10.7 & 29.8$\pm0.2$      \\
 0.790 & 43 & 112.6$\pm$59.0 & 178.8$\pm$93.5 & 111.6 &  111.6 &  12.7 &  16.5   &  -16.9  & 1.5$^{+0.3}_{-0.3}$ & 40.9$\pm0.2$   &  10.8 & 30.2$\pm0.2$      \\
 0.973 & 32 &  35.4$\pm$12.9 & 43.5$\pm$17.8  & 71.7  &  71.6  &  5.0  &  5.1    &  -17.1  & 1.2$^{+1.2}_{-0.1}$ & 40.9$\pm0.2$   &  10.8 & 30.1$\pm0.1$      \\
 1.124 & 12 &   6.1$\pm$4.8  & 9.4$\pm$6.8    & 34.7  &  34.7  &  1.3  &  1.5    &  $<$-17.2  & ...            & $<$40.9        &  10.9 & $<$30.1            \\
\hline
& & & & & \multicolumn{3}{|c|}{Faded Sample}  & & & & & \\
\hline                                                                                                                                                                                                               
 0.284 & 13 &  29.3$\pm$23.2 & 32.0$\pm$24.3  & 6.2   &  6.4   &  15.8 &  13.3   &  -16.5  &  1.0$^{+0.2}_{-0.1}$ & 40.3$\pm0.3$   &  10.7 & 29.7$\pm0.3$     \\ 
 0.518 & 11 &  35.1$\pm$17.6 & 51.9$\pm$24.5  & 17.6  &  18.2  &  10.9 &  13.0   &  -16.8  &  1.33$^{+0.03}_{-0.45}$ & 40.5$\pm0.2$   &  10.7 & 29.9$^{+0.3}_{-0.2}$   \\
 0.632 & 13 &  49.4$\pm$34.9 & 79.4$\pm$48.4  & 37.3  &  37.7  &  9.8  &  12.6   &  -17.1  &  1.5$^{+0.2}_{-0.6}$ & 40.4$\pm0.3$   &  10.6 & 29.8$\pm0.2$    \\
 0.767 & 15 & 105.8$\pm$54.9 & 151.0$\pm$85.2 & 41.7  &  41.7  &  19.4 &  22.3   &  -16.9  &  1.3$^{+0.1}_{-0.4}$ & 40.8$\pm0.2$   &  10.6 & 30.2$\pm0.3$   \\
 0.980 &  5 &  16.5$\pm$8.5  & 18.4$\pm$9.6   & 13.4  &  13.3  &  5.4  &  4.8    &  -17.4  &  1.2$^{+0.5}_{-0.4}$ & 40.6$\pm0.2$   &  10.7 & 30.0$\pm0.3$   \\
 1.119 &  3 &   2.6$\pm$3.1  & 2.8$\pm$1.4    & 9.6   &  9.6   &  1.0  &  0.8    &  $<$-17.5  & ...             & $<$40.7       &  10.6 &  $<$30.0        \\
\hline
\end{tabular}

\label{tab:SB1stack}
}
\end{center} 
\begin{flushleft}
{\sc Notes}: Column (1): Mean redshift of bin. Column~(2): Number of sources in
stacking bin.  Columns~(3)--(4): Net counts for SB1 and SB2. Columns~(5)--(6):
Exposure times in Ms for SB1 and SB2. Columns~(7)--(8): Stacked signal-to-noise
ratio for SB1 and SB2. Column~(9): Logarithm of the stacked SB1 flux (ergs
s$^{-1}$~cm$^{-2}$). Column~(10): X-ray temperature (keV). Column~(11):
Logarithm of the stacked \hbox{0.5--2~keV} luminosity $L_{\rm X}$ (ergs
s$^{-1}$) measured using the SB1 flux and an assumed Raymond-Smith plasma SED
with 1.5~keV temperature. Column~(12): Logarithm of the $B$-band luminosity
(for the faded sample, we list $L_{B, z= 0}$). Column~(13) Logarithm of the
0.5--2~keV to $B$-band luminosity ratio (for the faded sample, we list $L_{\rm
X}/L_{B, z=0}$).  \\ 
\end{flushleft}
\end{table*}

\begin{figure*}
\begin{center}{
\psfig{figure=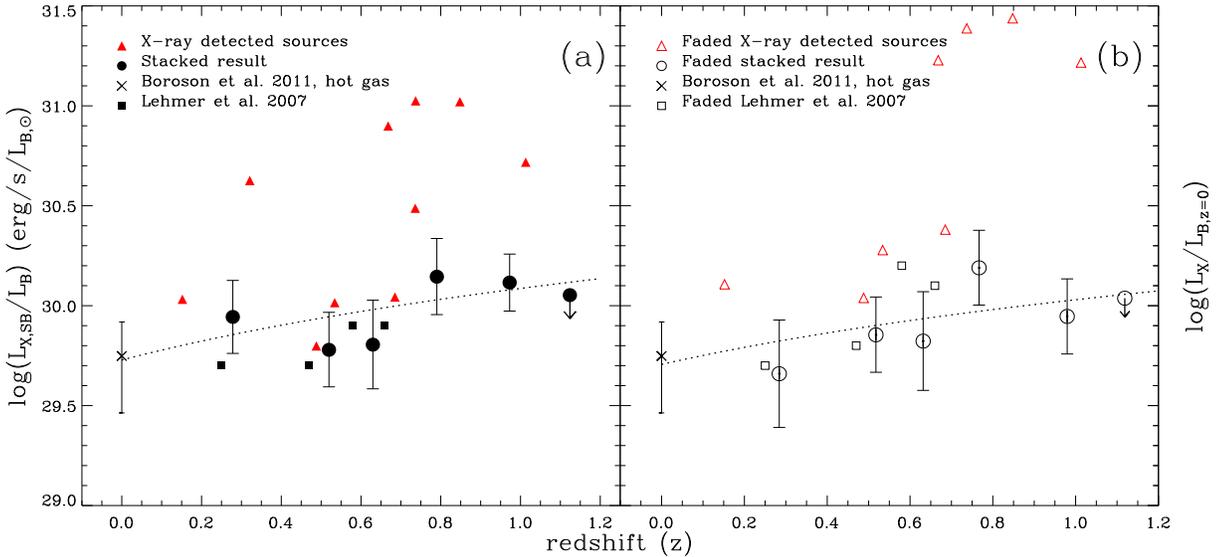,width=0.9\linewidth}
\caption{(a): The evolution of the soft X-ray properties of hot-gas
  dominated ETGs in terms of $B$-band optical luminosity, including
  passive X-ray detected galaxies, passive X-ray undetected galaxies
  and radio AGN. All galaxies with individual detections in the SB
  band are plotted as triangles.  The circles represent the stacked
  results in six redshift bins with only an upper limit in the highest
  redshift bin. {\it There is very little evolution of
    L$_{X,SB1}$/L$_B$ in these hot gas dominated galaxies, in
    particular given the very short expected cooling time of hot
    gas}. To minimise the contribution from LMXBs we stack in SB1
  (0.5-1keV) but we use the SED of a 1.5keV plasma to convert this to
  SB (0.5-2keV) in order to more easily compare the data to previous
  studies. $z=0$ points are shown for the hot gas contribution as
  determined from a subset of the ETG sample of \citet{Boroson11}. The
  simple two-parameter fit for $\log L_{\rm X}/L_{\rm B} = A + B \log
  (1+z)$ is shown as a dotted curve on both panels (a) and (b), with
  best-fit values of [$A$,$B$] = [$29.73\pm0.14$, $1.19\pm0.68$] and
  [$29.71\pm0.14$, $1.07\pm0.68$] for the main and faded samples,
  respectively. (b): Represents the faded sample, showing that the
  fading of L$_{\rm B}$ with cosmic time is a fundamental effect and
  acts to slightly decrease the observed L$_{\rm X,SB}/$L$_{\rm B}$
  ratio with increasing redshift. The errors on the luminosity are
  derived from propagating the bootstrapped errors on the net counts
  and systematic errors on the conversion between count-rate and flux
  using the 1.5keV Raymond-Smith plasma SED. } \label{fig:stackplot}
}
\end{center}
\end{figure*}

\subsection{X-ray Stacking Results}
\label{section:stack_results}

The results of stacking the X-ray data of the sample, are summarised
in Table~\ref{tab:SB1stack}. In Figures~3a and 3b, we have overlaid
our stacking results, demonstrating that our galaxies have SB2/SB1
count-rate ratios consistent with our adopted 1.5~keV Raymond-Smith
plasma SED (which we use to convert count-rates to fluxes and
luminosities) and $L_{\rm X}$ versus $L_{B,z=0}$ values consistent
with local hot gas dominated ETGs.  In Figures~7a and 7b, we display
the rest-frame \hbox{0.5--2~keV} luminosity (computed following
equation~2) per $B$-band luminosity $L_{\rm X}$/$L_B$ versus redshift
for our main and faded samples, respectively ({\it circles\/}).  The
10 individually X-ray detected, hot gas dominated ETGs are shown as
triangles in Figures~\ref{fig:stackplot}a and 7b (only eight are shown on Fig.~7b as only eight fulfilled the constraints of the stacking for the faded sample). Only 10 out of the
12 galaxies shown in Table~\ref{tab:XRayMatches} are included in the stacking as, for
stacking, we add the limitation that all galaxies in the E-CDF-S must
have $z<0.6$. As would be expected from an X-ray selected subset,
these sources generally have higher values of $L_{\rm X}/L_B$.  For
comparison, we have also plotted the mean $L_{\rm X}/L_B$ values
obtained by Lehmer \etal\ (2007) for ETGs with $L_B >
10^{10}$~$L_{B,\odot}$ ({\it squares\/}).

To constrain evolution to $z=0$, we take the \cite{Boroson11} sample
of 30 nearby ETGs and select only those 14 with $L_B
=$~(3--30)~$\times 10^{10} L_{\rm B,\odot}$.  We convert their
0.3--8~keV luminosities to 0.5--2~keV luminosities using our adopted
1.5~keV Raymond-Smith plasma SED and find a mean value of $\log L_{\rm
  X}/L_B = $~29.7~$\pm$~0.2 ({\it crosses} in Figs.~7a and 7b).  The
combination of the Boroson \etal\ (2011) mean $L_{\rm X}/L_B$ and our
stacking results indicates that there is little apparent evolution
in $L_{\rm X}/L_{\rm B}$ for these optically luminous ETGs. However, a
Spearman's $\rho$ test reveals that the quantity $L_{\rm X}/L_{\rm B}$
is correlated with $z$ at the 92\% and 96\% probability level for the
main and faded samples respectively.  To constrain the allowable
redshift evolution of $L_{\rm X}/L_{\rm B}$, we fit a simple two
parameter model to the data $\log L_{\rm X}/L_{\rm B} = A + B \log
(1+z)$ and find best-fit values of [$A$,$B$] = [$29.73\pm0.14$,
  $1.19\pm0.68$] and [$29.71\pm0.14$, $1.07\pm0.68$] for the main and
faded samples, respectively.  These values indicate mild evolution in
the X-ray activity of luminous ETGs and are consistent with those of
\cite{Lehmer07}. Using this model, we find that at z=1.2, ETGs are
$\sim2.4\pm0.9$ times or $\sim2.2\pm0.9$ times (for the main and faded
sample respectively) more X-ray luminous (per unit L$_{B}$) than at
z=0, which suggests only modest evolution. Our best-fit relations have
been highlighted in Figures~7a and 7b as dashed curves.

Since the stacked X-ray properties (i.e., SB2/SB1 band ratio and
$L_{\rm X}$ versus $L_B$) are consistent with those expected from hot
gas dominated ETGs, with little expected contributions from LMXBs, we
can use the \xray\ luminosity versus redshift diagram for our stacked
samples as a direct tracer of the hot gas cooling history for massive
ETGs with $L_{B} = (3-30) \times 10^{10} L_{B, \odot}$.  The observed
mild decline in \xray\ luminosity per unit $B$-band luminosity and
roughly constant \xray\ gas temperature for massive ETGs over the last
$\approx$8.4~Gyr of cosmic history suggest that, on average, the gas
is being kept hot.  We expect that many complex processes are
contributing to the evolution of the gas including radiative cooling,
periodic AGN heating and outflows, replenishment from stellar winds
and supernovae, interactions and sloshing, and intergalactic medium
and poor group inflow (e.g. \citealt{Tabor93}, \citealt{Best06},
\citealt{Faber76} and \citealt{Brighenti99}).  The detailed influences
that each of these processes has on the gas are difficult to quantify,
particularly without a strong idea of the environment in which each
galaxy resides. However, we know that most of our sources reside
within small groups and clusters, therefore, processes such as
intergalactic medium and poor group infall may be important. One of
the goals of this paper is to test whether AGN feedback from
mechanical feedback can provide enough energy to keep the gas hot and
counter the observed cooling over the long baseline of cosmic time
spanned by our observations.

In the next section, we discuss the viability of AGN feedback heating of the
gas by directly measuring the history of radio AGN events in our galaxy
population and computing the mechanical energy available from these events.

\section{Discussion}

\subsection{The Hot Gas Cooling and Mechanical Heating Energy Budgets}
\label{sec:energy}

The above \xray\ stacking results indicate that the \xray\ power output from
hot gas in the massive ETG population remains well regulated across a large
fraction of cosmic history (since $z \sim 1$).  To determine whether the
heating from AGNs is sufficient to keep the gas hot, we estimated the
mechanical power input from AGNs and the radiative cooling power from the hot
gas.  As discussed in $\S$~4.2, the history of gas cooling power can be
directly inferred from our \xray\ stacking results; the gas cooling power, can
be expressed as
\begin{equation}
L_{\rm cool} =  C_{\rm bol} L_{\rm X} \approx C_{\rm bol} L_{B} 10^A (1 + z)^B,
\end{equation}
where $A = 29.73 \pm 0.14$ and $B = 1.19 \pm 0.68$ were computed in
$\S$4.2, $L_{B} \approx 6.3 \times 10^{10}$~$L_{B,\odot}$ is the mean
value of $L_{B}$, and $C_{\rm bol} \approx 1.8$ is the bolometric
correction for a hot gas SED with 1.5~keV temperature (see $\S$4.2
above).  In Fig.~8, we plot the mean cooling history ({\it filled
  circles} and {\it dashed curve} for stacked values and best-fit
model, respectively), since $z \approx 1.2$.

To estimate the energy input from radio AGNs over the last
$\approx$8.4~Gyr of cosmic history, we began by measuring the radio
AGN fraction as a function of radio luminosity (a proxy for mechanical
heating) and redshift.  By making the assumption that all galaxies
will go through multiple AGN active phases, we can use the
radio-luminosity and redshift dependent AGN fraction as a proxy for
the typical AGN duty cycle history for galaxies in our sample.

To establish a baseline local ($z \approx 0$) measurement of the ETG
radio AGN fraction, we used the B05 sample of radio-loud AGN from the
SDSS survey, which included both early-type and late-type galaxies.
For the sake of comparing these data with our ETG sample, we selected
galaxies in the B05 sample with elliptical-like concentration indices
$C > 2.6$ \citep{Strateva01}.  The concentration index is defined as
$C = r_{90}/r_{50}$, where $r_{90}$ and $r_{50}$ are radii containing
90\% and 50\% of the optical light respectively.  By applying the
  flux density limit of 5~mJy, we limit the B05 sample to a lower
  radio luminosity limit of $\approx$$10^{23}$~W~Hz$^{-1}$, which
  corresponds to a maximum redshift of $z=0.1$.  To measure the AGN
  fraction for galaxies in the distant Universe, we used the sample of
  distant ETGs presented in this paper.  Using $L_{\rm 1.4~GHz} =
  10^{23}$~W~Hz$^{-1}$, the luminosity limit used for the B05 data, we
  determined that the corresponding \hbox{CDF-N} and \hbox{E-CDF-S}
  radio flux limits (see $\S$3.3) allow us to study similar AGNs out
  to $z \approx 1$ and 0.85, respectively.  We calculated the AGN
  fraction for both the local B05 local galaxies and our distant
  galaxies in three bins of radio luminosity (in even logarithmic
  luminosity intervals) in the range of $L_{\rm 1.4~GHz} \sim (1-100)
  \times 10^{23}$~W~Hz$^{-1}$, each bin with a different allowed
  redshift range due to the flux limits. These bins in luminosity and
  redshift then result in a total of 2642 elliptical galaxies
  containing a radio AGN from the B05 sample and 13 radio AGN from our
  sample. The AGN fraction was computed in each bin, for both the B05
  sample and our sample, by taking the total number of radio AGN in a
  particular luminosity range and dividing it by the total number of
  galaxies in which an AGN with a luminosity lying within that range
  could have been detected if present. We estimate 1$\sigma$ errors on
  the AGN fractions following \cite{Gehrels86}.  By comparing the
  radio AGN fraction at the mean redshift $z=z_{\rm mean}$ of our
  distant galaxy sample with that of the B05 $z \approx 0$ sample, we
  can estimate the evolution of the duty cycle of the AGN outbursts in
  each radio luminosity bin.  The time-dependent radio AGN fraction
  for each bin of radio-luminosity was computed following
\begin{equation}
f_{AGN}(t,L_{{\rm 1.4~GHz},i})=\frac{f_{ {\rm AGN},z=z_{{\rm zmean},i}}-f_{
{\rm AGN},z \approx 0, i}}{\Delta t_i} t+f_{{\rm AGN},z\approx0, i}, 
\end{equation} 
where $\Delta t_i$ is the difference in the mean lookback time between $z =
z_{{\rm mean},i}$ and $z = 0$ (i.e., our AGN fraction and that of B05) in a
particular bin of mean radio luminosity $L_{{\rm 1.4~GHz},i}$.  

Several studies have now shown that the radio power output $L_{\rm 1..4~GHz}$
from AGNs within nearby giant elliptical (gE) and cluster central galaxies
correlates with the inferred mechanical power $L_{\rm mech}$ that is needed to
inflate the cavities within hot X-ray halos (e.g., \citealt{Birzan04};
\citealt{Birzan08}; \citealt{Cavag10}; \citealt{Osul11}).  Until recently, these
relations have been calibrated using the cores of cooling clusters, and may not
be appropriate for the massive early-type galaxies studied here. \cite{Cavag10}
have added a sample of 21 gE galaxies and have shown that, as long as the radio
structures are confined to the hot \xray\ emitting gas region, gE galaxies
provide a natural extension to the $L_{\rm mech}$--$L_{\rm 1.4~GHz}$
correlation at low $L_{\rm mech}$.  However, as noted by \cite{Cavag10}, gE
galaxies and FRI sources in group environments (e.g., \citealt{Cros08}) tend to
have $L_{\rm mech}/L_{\rm 1.4~GHz}$ ratios much lower than the correlation
derived including clusters.  Since our galaxies are expected to be gE and group
central galaxies, we made use of $L_{\rm mech}$ and $L_{\rm 1.4~GHz}$ values
for the sample of 21 gEs from \cite{Cavag10} to derive the $L_{\rm
mech}$--$L_{\rm 1.4~GHz}$ correlation for these sources.  Figure~8 shows the 21
gEs from \cite{Cavag10}.  We find that the best-fit relation from
\cite{Cavag10} ({\it dashed line} in Fig.~8), which includes radio galaxies at
the centers of galaxies, overpredicts the $L_{\rm mech}/L_{\rm 1.4~GHz}$ ratios
for AGNs with $L_{\rm 1.4~GHz} \simgt 10^{22}$~W~Hz$^{-1}$.  Using these data,
we derived the following relation, which is applicable to gE galaxies:
\begin{equation}
L_{\rm mech} \approx 3.36 \times 10^{35}  \left(\frac{L_{\rm 1.4~GHz}}{10^{24}
{\rm W~Hz^{-1}}}\right)^{0.11} {\rm W}.
\end{equation}
Our best-fit relation is plotted in Figure~8 as a solid line.

\begin{figure}
\begin{center}
\psfig{figure=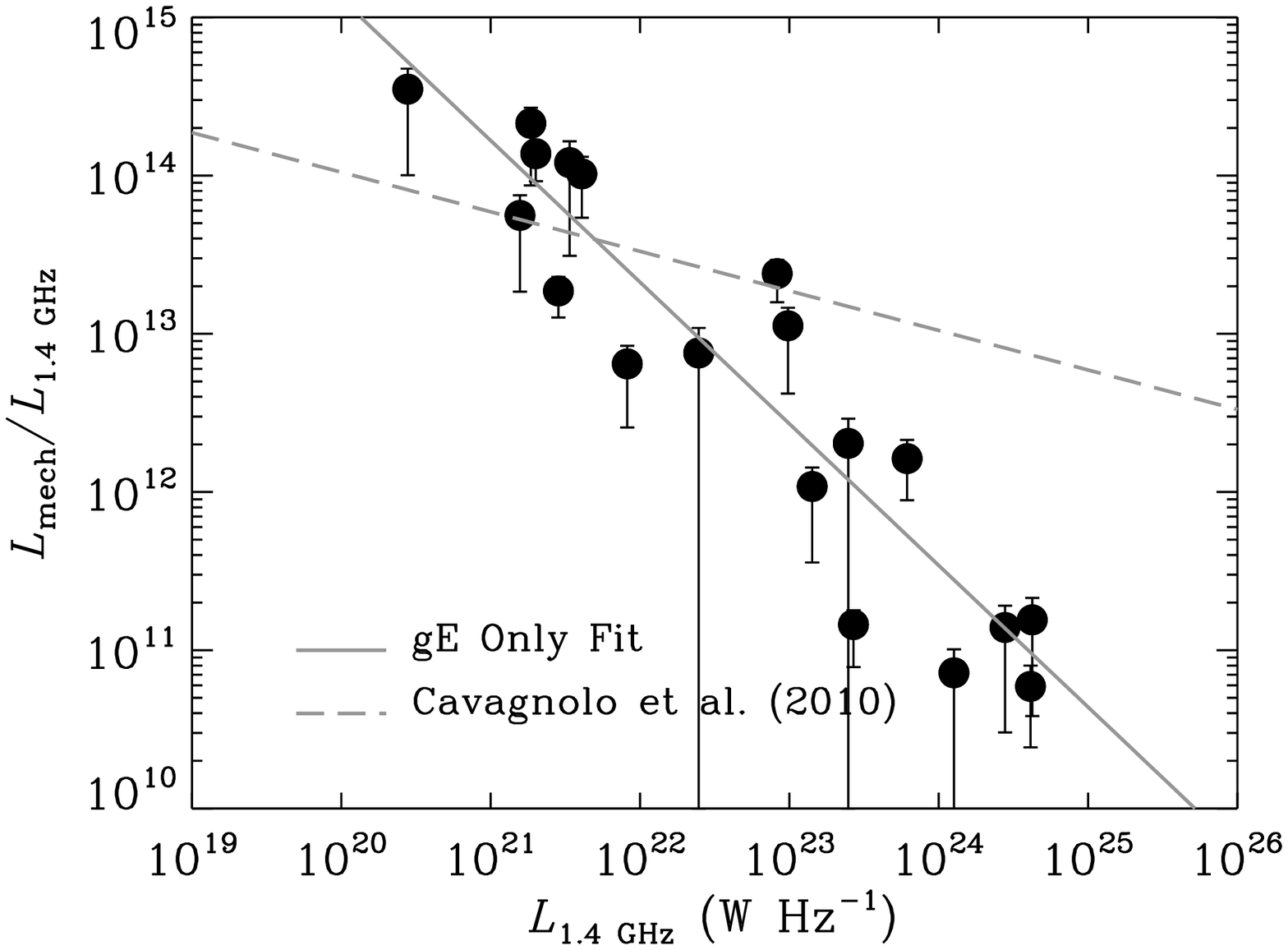,width=3.5in}
\caption{
Ratio of mechanical power to 1.4~GHz radio luminosity ($L_{\rm mech}/L_{\rm
1.4~GHz}$) versus $L_{\rm 1.4~GHz}$ for the 21 giant elliptical (gE) galaxies
studied by Cavagnolo \etal\ (2010; {\it filled circles\/}).  The Cavagnolo
\etal\ (2010) relation, which includes gE galaxies and radio galaxies at the
centers of cooling clusters (from \citealt{Birzan08}), has been shown as a
dashed line.  Our best-fit relation for gE only galaxies, presented in
equation~6, is shown as a solid line.}
\label{fig:relation}
\end{center}
\end{figure}

Using equations~(5) and (6), we then estimate the average mechanical feedback
power per galaxy over the last $\approx$8.4~Gyr of cosmic history considering
all radio AGNs in the range of $L_{\rm 1.4~GHz} \sim$~(1--100)~$\times
10^{23}$~W~Hz$^{-1}$ via the following summation:
\begin{equation}
L_{\rm heating}= \sum_i f_{AGN}(t, L_{{\rm 1.4~GHz},i}) \; L_{\rm mech, i} 
\end{equation} 
In Fig.~9, we show the mean heating luminosity and 1$\sigma$ errors ({\it solid
curve} with {\it shaded envelope\/}) derived following equation~7.  From
Fig.~9, we see that on average there appears to be more than sufficient input
mechanical energy from radio AGN events to balance the hot gas radiative
cooling.  From the five stacked bins where we obtain \xray\ detections, we
estimate on average $L_{\rm heating}/L_{\rm cool} \approx 2.4^{+0.9}_{-0.5}$.
This result is broadly in agreement with that found for local elliptical
galaxies of comparable mass where the mechanical power has been measured using
\xray\ cavities (e.g., \citealt{Nulsen07}).  \cite{Nulsen07} estimate that the
total cavity heating can be anywhere between 0.25 and 3 times the total gas
cooling if 1~pV of heating is assumed per cavity; however, the enthalpy of the
cavity and therefore the total heating may be much higher.  Stott et al.
(submitted) find a trend in groups and clusters indicating the ratio of
intra-cluster medium (ICM) AGN heating (from the brightest cluster galaxy) to
ICM cooling increases with decreasing halo mass.  For halo masses
$<5\times10^{12}$M$_{\odot}$ the heating can exceed the cooling.  We have found
that most of our galaxies are likely to live in small group environments.
Therefore extrapolating this relation to the expected halo masses of the
galaxies in our sample ($\approx$[1--10]~$\times10^{12} M_\odot$ for $L_B
\approx$~[3--30]~$\times 10^{10}$~$L_{B,\odot}$; \citealt{Vale04}) would
similarly imply that the mechanical heating would likely exceed the radiative
cooling.

We note that our heating calculation is based on duty cycle histories derived
primarily from the $<$20 distant radio AGNs in
our sample and is based on the assumption that each galaxy will have
many radio outbursts that span the full range of radio luminosities
studied here.  We therefore expect these calculations will have
significant uncertainties that we cannot determine.  In the next
section, we estimate the global ETG hot gas cooling and radio AGN
heating power as a function of redshift.

\subsection{Cosmic Evolution of Global Heating and Cooling Density}
\label{section:heat_models}

Using a large sample of radio-loud AGNs in the local Universe, Best
\etal\ (2006) computed the radio-luminosity and black-hole mass
dependent AGN fraction of nearby galaxies.  Their data show that the
population-averaged mechanical power (probed by 1.4~GHz power)
produced by these AGN events increases with black-hole mass (and also
$B$-band luminosity) and balances well the radiative power output from
\xray\ cooling of the hot gas (see their Fig.~2).  Their analyses
further revealed that relatively low luminosity radio AGN ($\log
L_{\rm 1.4~GHz}/({\rm W~Hz^{-1}}) \approx$~22--25) are likely to
provide the majority of the mechanical feedback power for the
population as a whole.  They estimated that in the local universe, the
mean mechanical power output density from mechanical heating from
radio AGNs with $L_{\rm 1.4~GHz} > 10^{22} {\rm W~Hz^{-1}}$ is
$\approx$$4 \times 10^{31}$~W~Mpc$^{-3}$.

\begin{figure}
\begin{center}
\psfig{figure=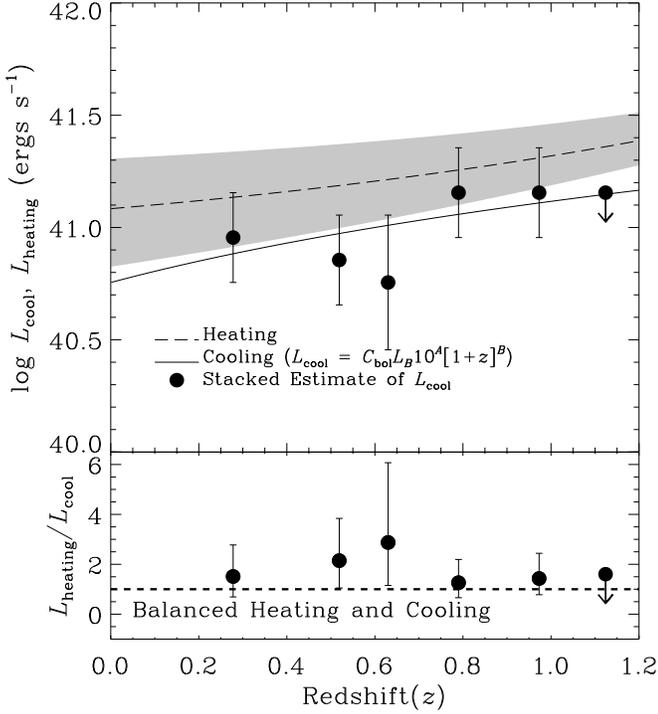,width=3.5in}
\caption{
({\it top panel\/}) Mean radiative cooling power $L_{\rm
cool}$ and mean mechanical heating power $L_{\rm heating}$ versus redshift.
The filled circles and 1$\sigma$ error bars show the
bolometrically corrected estimates of $L_{\rm cool}$ (see Table~\ref{tab:SB1stack}) and the
solid curve shows our best-fit model.  The {\it long-dashed curve} with shaded
region represents our best estimate of the mean heating luminosity as presented
in equation~7.  These measurements show that for the early-type galaxies in our
sample, there is more than enough energy available from radio AGN heating to
keep the gas from cooling.
({\it bottom panel\/}) Ratio of mean heating to radiative cooling luminosity
versus redshift.  We find that the average heating power is $\approx$1.4--2.6
times larger than the radiative hot gas cooling power.}  
\label{fig:AGN_cooling}
\end{center}
\end{figure}

\begin{figure}
\begin{center}
\psfig{figure=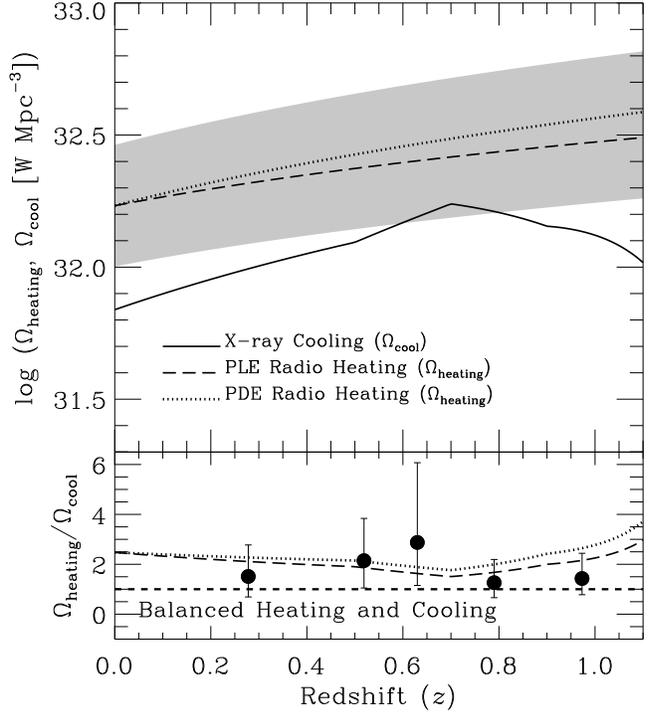,width=3.5in}
\caption{
({\it top panel\/}) Cosmic history of mechanical heating $\Omega_{\rm heating}$
and radiative cooling $\Omega_{\rm cool}$ of hot gas since $z \approx 1.1$.
The mechanical luminosity density versus redshift for radio AGNs, as computed
by \citet{Smolcic09}, for pure-luminosity evolution and pure density evolution
are indicated as dashed and dotted curves, respectively.  The shaded envelope
provides the estimated uncertainties between the $L_{\rm 1.4~GHz}$ and $L_{\rm
mech}$ correlation.  The cooling history of hot gas in ETGs is
indicated as a solid curve (see $\S$5.2 for details).
({\it bottom panel\/}) Ratio of mechanical heating to radiative cooling versus
redshift.  For comparison, we have plotted our estimates of the mean
heating-to-cooling luminosity ratios ($L_{\rm heating}/L_{\rm cool}$ as
provided in the bottom panel of Fig.~9 ({\it filled circles\/}).  This
relation shows that there is plenty of mechanical energy potential available to
keep the gas hot over the last $\approx$8.4~Gyr.  
}
\label{fig:AGN_cooling}
\end{center}
\end{figure}

Due to the relatively small number of radio AGNs found in our survey, it is not
feasible to calculate the evolution of the radio and mechanical luminosity
density of the Universe.  However, the evolution of the radio AGN luminosity
function has recently been measured out to $z \approx 1.3$ using the VLA-COSMOS
survey \citep{Schin07} to relatively faint luminosity levels ($L_{\rm 1.4~GHz}
\simgt 10^{21}$--$10^{23}$~${\rm W~Hz^{-1}}$; \citealt{Smolcic09}).  By
converting radio luminosity into mechanical luminosity, \cite{Smolcic09}
integrated their luminosity functions to determine the estimated mechanical
power density of the Universe out to $z \approx 1.3$.  In Fig.~10, we show
the expected mechanical feedback power density evolution, based on the
\cite{Smolcic09} radio luminosity function and equation~6 for both
pure-luminosity density evolution ($L_* \propto [1+z]^{0.8}$) and pure-density
evolution ($\Phi_* \propto [1+z]^{1.1}$), the best-fit parameterisations for
the evolution of the 1.4~GHz luminosity function.

As shown in $\S$~4, our \xray\ stacking measurements can be described on
average as $L_{\rm cool, mod} \approx C_{\rm bol} 10^A (1+z)^B L_B$ (with
$C_{\rm bol} = 1.8$, $A = 29.73 \pm 0.14$, $B = 1.19 \pm 0.68$, and $L_B
\approx 6.3 \times 10^{10}$~$L_{B,\odot}$) for the ETG population.  Using this
scaling relation and the observed evolution of the ETG $B$-band luminosity
function $\varphi_B$ from Faber \etal\ (2007; see their Table~4), we can
compute the expected volume-averaged cooling luminosity density.  In this
exercise, we assumed a constant intrinsic scatter of $\sigma \approx$~1~dex for
the $L_{\rm X}/L_B$ ratio (Boroson \etal\ 2011) and transformed the $B$-band
luminosity function into a \xray\ gas cooling luminosity function using the
following transformation:
\begin{eqnarray}
\varphi_{\rm X}(\log L_{\rm X}, z) = \int_{-\infty}^{\infty} \varphi_B(\log
L_B, z) P(\log L_{\rm X} \vert \log L_B) \; d\log L_B \nonumber \\
P(\log L_{\rm X} \vert \log L_B) = \frac{1}{\sqrt{2\pi} \sigma} \exp \left[
{-\frac{(\log L_{\rm cool,mod} - \log L_{\rm X})^2}{2 \sigma^2}} \right].
\end{eqnarray}
The total redshift-dependent cooling density $\Omega_{\rm cool}(z)$ of the
Universe can therefore be computed following
\begin{equation}
\Omega_{\rm cool}(z) = \int_0^{\infty} \varphi_{\rm X}(\log L_{\rm X},z) \;
L_{\rm X} \; d\log L_X.
\end{equation}
In Fig.~10, we show the resulting $\Omega_{\rm cool}(z)$ versus
redshift ({\it solid curve\/}).  Our analyses show that the estimated
mechanical power provided by radio AGN activity is a factor of
  \hbox{$\approx$1.5--3.5} times larger than the radiative cooling
power (see bottom panel of Fig.~10), and the shape of the heating and
cooling histories appear to be in good overall agreement.  For
comparison, we plot the mean values of $L_{\rm heating}/L_{\rm cool}$
as measured in $\S$~5.1 and Figure~9 ({\it filled circles\/}), which
are in agreement with the global heating-to-cooling estimates obtained
here.

The combination of the approaches for estimating global and mean
galaxy heating and cooling taken here and in $\S$~5.1, respectively,
indicate that mechanical heating exceeds that of the radiative gas
cooling for early-type galaxies with $L_B \approx$~(3--30)~$\times
10^{10}$~$L_{B,\odot}$.  These computations are based on the
assumption that the radio luminosity provides a direct proxy for
mechanical power, which scales for our early-type galaxies in the same
way as that measured for local gEs (i.e., based on data from
  \citealt{Cavag10}). Indeed, some studies have suggested that AGN
heating in less massive systems, like those studied here, may have
different heating cycles and mechanical efficiencies (\citealt{G11}).
Future studies that characterize how the radio and mechanical power
are related in galaxies like those studied here would be needed to
exclude the possibility that the excess of mechanical power compared
with cooling power (as observed here) is due to the calibration.

\section{Summary and Future Work}  

The X-ray and multiwavelength properties of a sample of 393 massive
ETGs in the $\Chandra$ Deep Field surveys have been studied, in order
to constrain the radiative cooling and mechanical feedback heating
history of hot gas in these galaxies.  We detected 55 of the galaxies
in our sample in the \xray\ bandpass, and using the \xray\ and
multiwavelength properties of these sources, we find that 12 of these
systems are likely to be dominated by \xray\ emission from hot gas.
To measure the evolution of the average ETG \xray\ power output, and
thus hot gas cooling, we stacked the \hbox{0.5--1~keV} emission of the
\xray\ undetected and detected ``normal'' galaxy population in
redshift bins.

We find that the average rest-frame \hbox{0.5--2~keV} luminosity per unit
$B$-band luminosity ($L_{\rm X}/L_B$) has changed very little since $z \approx
1.2$ and is consistent with $\propto (1+z)^{1.1\pm0.7}$ evolution.  This suggests
that the population average hot gas power output is well regulated over
timescales of $\approx$8~Gyr; much longer than the typical cooling timescale of
the hot gas ($\approx$0.1--1~Gyr).  We hypothesize that mechanical heating from
radio luminous AGNs in these galaxies is likely to play a significant role in
keeping the gas hot, and we compare the implied gas cooling from our stacking
analyses with radio-AGN based estimates of the heating.

We find that if local relations between radio luminosity and
mechanical power hold at high redshifts, then the observed
radio-luminosity dependent AGN duty cycle suggests that there would be
more than sufficent  (factor 1.4--2.6 times) mechanical energy
needed to counter the inferred cooling energy loss.  Similarly, we
find that the evolution of the mechanical power density of the
Universe from radio AGNs increases only mildly with redshift and
  remains a factor of $\approx$1.5--3.5 times higher than the
radiative hot gas cooling power density of ETGs in the Universe.
These results are concordant with previous lower redshift studies (e.g
\citealt{Best05} and \citealt{Lehmer07}) and with theoretical feedback
models such as \cite{Churazov05}, \cite{Croton06} \cite{Bower08} and
\cite{Bower06} where feedback from radio AGN maintains the balance
between heating and cooling rates of hot interstellar gas in massive
ETGs.\\

Understanding the evolution of both the X-ray and radio properties of optically
luminous ETGs could be improved in future work by: (1) gaining a better
understanding of the environments of these sources in order to better constrain
the likely contribution of e.g. gas infall to the evolution of the X-ray
properties and probe the influence of environment on the balance between gas
cooling and heating in galaxies. This could be achieved by measuring
spectroscopic redshifts for the whole sample, as there is currently significant
uncertainty in photometric redshifts.  (2) Conductng deeper X-ray observations
could provide stronger constraints on the evolution of the hot gas. We focus
our observations in the soft band in which the background is lowest, therefore
doubling the Chandra exposure time to 8Ms could provide a factor of 1.4--1.6
improvement in the sensitivity resulting in the faintest detectable sources
having soft band fluxes of $6.0\times10^{-18}$erg cm$^{-2}$ s$^{-1}$
\citep{Xue11} and improved statistics on the average X-ray emission from the
ETG population.

\section*{acknowledgements}

We thank the anonymous referee for their helpful comments. ALRD
acknowledges an STFC studentship. We would like to thank Ian Smail for
useful comments and feedback on this work. We also thank Philip Best
for providing us with his sample for comparing to our work and Laura
B\^{i}rzan for useful advice. BDL acknowledges financial support from
the Einstein Fellowship program. WNB and YQX thank CXC grant
SP1-12007A. DMA acknowledges financial support from STFC.

\bibliography{ref.bib}
\bibliographystyle{mn2e}
\bsp

\label{lastpage}
\end{document}